\documentclass[aps,reprint,superscriptaddress]{revtex4-1}
\usepackage{graphicx,xcolor,soul}
\pdfoutput=1
\usepackage{amsmath}
\usepackage{float}
\usepackage{amssymb}
\usepackage[colorlinks=true,citecolor=blue,breaklinks=true,%
urlcolor=blue,linkcolor=blue]{hyperref}

\newcommand{\vect}[1]{\boldsymbol{#1}}
\newcommand{\op}[1]{\hat{\boldsymbol{#1}}}

\newcommand{\px}{\tilde{p}_{x}}
\newcommand{\py}{\tilde{p}_{y}}

\begin{document}

\title{Tunable Fermi surface topology and Lifshitz transition in bilayer graphene}

\author{Anastasia Varlet}
\email[]{varleta@phys.ethz.ch}
\affiliation{Solid State Physics Laboratory, ETH Z\"{u}rich, 8093 Z\"{u}rich, Switzerland}

\author{Marcin Mucha-Kruczy\'nski}
\affiliation{Department of Physics, University of Bath, Claverton Down, Bath, BA2 7AY, UK}

\author{Dominik Bischoff}
\affiliation{Solid State Physics Laboratory, ETH Z\"{u}rich, 8093 Z\"{u}rich, Switzerland}

\author{Pauline Simonet}
\affiliation{Solid State Physics Laboratory, ETH Z\"{u}rich, 8093 Z\"{u}rich, Switzerland}

\author{Takashi Taniguchi}
\affiliation{Advanced Materials Laboratory, National Institute for Materials Science, 1-1 Namiki, Tsukuba 305-0044, Japan}

\author{Kenji Watanabe}
\affiliation{Advanced Materials Laboratory, National Institute for Materials Science, 1-1 Namiki, Tsukuba 305-0044, Japan}

\author{Vladimir I. Fal'ko}
\affiliation{Department of Physics, Lancaster University, Lancaster, LA1 4YB, United Kingdom}

\author{Thomas Ihn}
\affiliation{Solid State Physics Laboratory, ETH Z\"{u}rich, 8093 Z\"{u}rich, Switzerland}

\author{Klaus Ensslin}
\affiliation{Solid State Physics Laboratory, ETH Z\"{u}rich, 8093 Z\"{u}rich, Switzerland}

\date{\today}

\begin{abstract}
Bilayer graphene is a highly tunable material: not only can one tune the Fermi energy using standard gates, as in single-layer graphene, but the band structure can also be modified by external perturbations such as transverse electric fields or strain. We review the theoretical basics of the band structure of bilayer graphene and study the evolution of the band structure under the influence of these two external parameters. We highlight their key role concerning the ease to experimentally probe the presence of a Lifshitz transition, which consists in a change of Fermi contour topology as a function of energy close to the edges of the conduction and valence bands. Using a device geometry that allows the application of exceptionally high displacement fields, we then illustrate in detail the way to probe the topology changes experimentally using quantum Hall effect measurements in a gapped bilayer graphene system.
\end{abstract}

\maketitle

\section*{\label{sec:intro}Introduction}

The shape of the Fermi surface is crucial for understanding the electronic properties of metals \cite{abrikosov_book_1988}. As first noticed by Lifshitz \cite{lifshitz}, changes in the Fermi surface topology cause anomalous behavior of thermodynamic, transport and elastic properties of materials \cite{blanter_physrep_1994}. Intuitively, the simplest way to observe such an electronic topological transition, also known as a Lifshitz transition, is by tuning the Fermi level to the singular point in the band structure where the change of topology takes place. However, this usually requires considerable variations of the electron density that for bulk materials can only be achieved by the experimentally inconvenient preparation of numerous samples with different chemical compositions \cite{bruno_physrep_1994, yoshizumi_jpsj_2007, okamoto_prb_2010, sebastian_pnas_2010, norman_prb_2010, leboeuf_prb_2011}. Alternatively, the Fermi surface can be deformed by exposing the sample to extreme conditions like high pressures \cite{lifshitz, chu_prb_1970, godwal_prb_1998} or strong magnetic fields \cite{rourke_prl_2008, wosnitza_physb_2008}.

\begin{figure}
\includegraphics[width=\columnwidth]{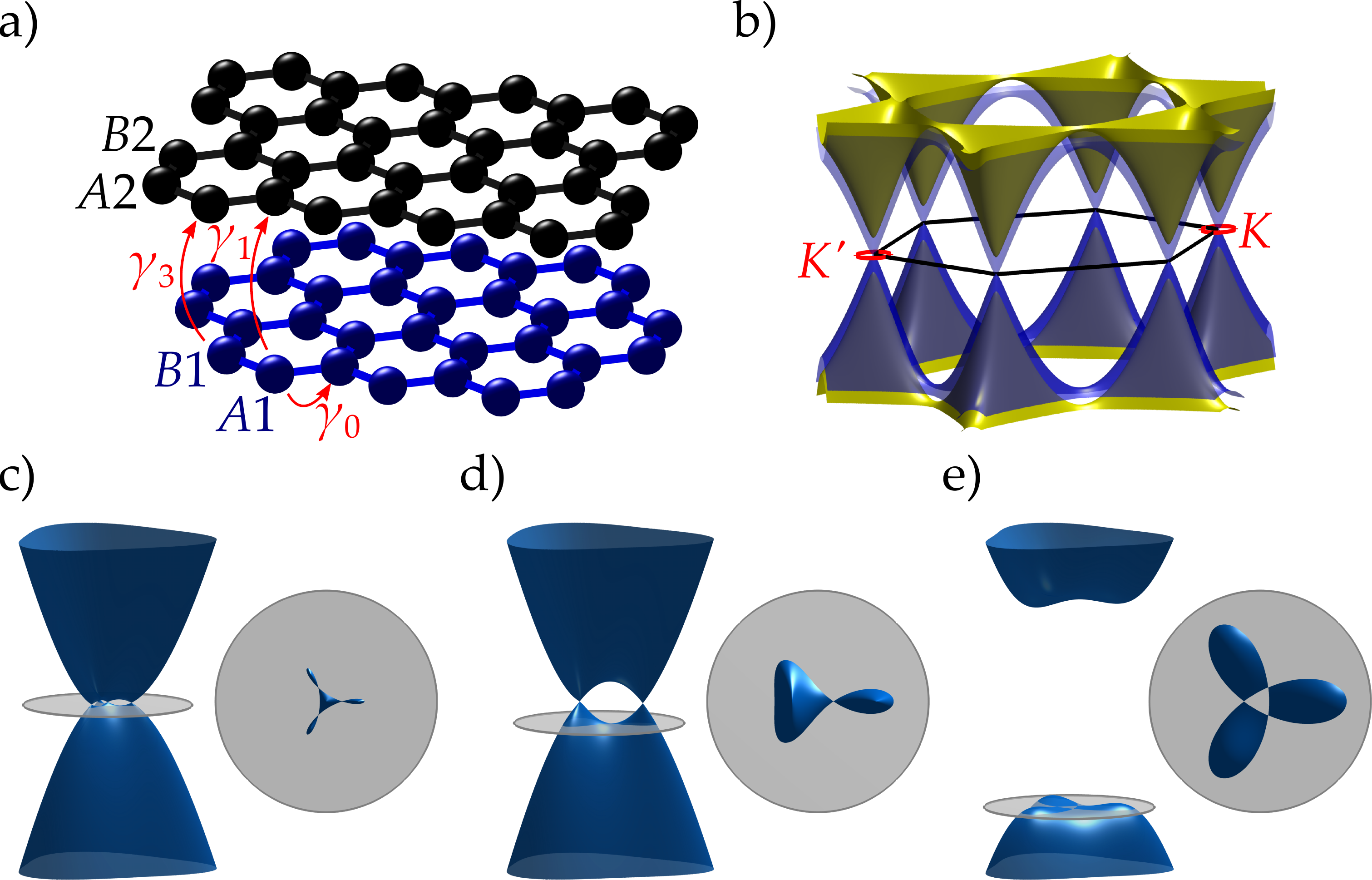}
\caption{(a) Schematic lattice structure of bilayer graphene. Also shown are all the electron hoppings mentioned in this work. (b) Electronic structure of bilayer graphene on the energy scale of 3~eV from the neutrality point (marked by the position of the black hexagon which shows bilayer graphene Brillouin zone). The blue (yellow) surfaces show the low-energy (split) bands. Red circles indicate the $K$ and $K'$ valleys. (c-e) Low-energy electronic structure at the $K$ valley of an (c) unperturbed, (d) homogeneously strained and e) gapped bilayer graphene. Grey discs cut through the dispersion at the energy of the Lifshitz transition and the corresponding cross-section is shown in the right of each of the panels.}
\label{fig:intro}
\end{figure}

In this article we discuss the theoretical prediction and an experimental study of the Lifshitz transition in bilayer graphene \cite{mccann_landau-level_2006}, two coupled layers of carbon atoms arranged in a hexagonal lattice [Fig.~\ref{fig:intro}(a)]. Its electronic dispersion relation contains saddle points, similar to those well known from graphite \cite{mcclure1957, inoue_jpsj_1962, williamson_ssc_1966, orlita_prl_2012}. However, as opposed to bulk graphite, external perturbations like strain \cite{mucha-kruczynski_prb_2011, mucha-kruczynski_ssc_2011} or electric fields \cite{varlet_anomalous_2014} can modify the topology of the electronic dispersion and change the energy of the Lifshitz transition which connects regions of different Fermi contour topologies [Fig.~\ref{fig:intro}(c-e)]. Moreover, due to the two-dimensional nature of bilayer graphene, its chemical potential and its topology can be tuned with electrostatic gates \cite{novoselov2004}, simplifying experimental studies of the Lifshitz transition.

		%%%%% Fig - lattices %%%%%
		%%%%% End of Fig - lattices %%%%%

We first discuss in detail the band structure of bilayer graphene in the absence and presence of an external quantizing magnetic field perpendicular to the carbon layers. We follow on with the analysis of the effects on the electronic spectrum of external parameters, such as homogeneous strain and electric fields applied across the layers. We then describe state-of-the-art bilayer graphene devices and the experimental signatures of the Lifshitz transition in this material.

\section{\label{sec:bands}Electronic band structure of neutral bilayer graphene}
\subsection{In the absence of the magnetic field}

Bilayer graphene, shown in Fig.~\ref{fig:intro}(a), is composed of two graphene \cite{geim_science_2009} layers, that is, two layers of hexagonally arrayed carbon atoms. The unit cell consists of four atoms which we label as $A1$, $B1$, $A2$ and $B2$, with $A$ and $B$ labeling the sublattice within the same layer and the index 1 (2) denoting the bottom (top) layer. In the present study, we focus on Bernal-stacked bilayer graphene, i.e. atom $A2$ lies on top of atom $B1$. For the electronic properties, it is enough to consider only the delocalized $\pi$-electrons, one per carbon atom. Within the tight-binding approach, the essential features of the electronic dispersion are captured by taking into account the in-plane nearest neighbor electron hopping $\gamma_{0} = \gamma_{A1B1} = \gamma_{A2B2} \approx 3$~eV, the vertical interlayer coupling $\gamma_{1} = \gamma_{A2B1} \approx 0.4$~eV and the skew interlayer hopping $\gamma_{3} = \gamma_{A1B2} \approx 0.4$~eV \cite{kuzmenko_prb_2009}. The resulting electronic bands are shown in Fig.~\ref{fig:intro}(b) within the hexagonal Brillouin zone containing two inequivalent corners $K$ and $K'$, also called valleys. Two of the bands, which are shown in blue in Fig.~\ref{fig:intro}(b) and which we refer to as low-energy ones, touch at the center of the valley at the energy which marks the neutrality point -- the position of the chemical potential in neutral bilayer graphene. The other two, called split bands and shown in yellow in Fig.~\ref{fig:intro}(b), are split off the neutrality point by the energy $\epsilon\approx\pm\gamma_{1}$. In the vicinity of the valley, the electronic band structure can be described by a four-band Hamiltonian \cite{mccann_landau-level_2006} written in the basis of sublattice Bloch states $(\phi_{A1},\phi_{B2},\phi_{A2},\phi_{B1})^{T}$ in valley $K$ and $(\phi_{B2},\phi_{A1},\phi_{B1},\phi_{A2})^{T}$ in valley $K'$,
\begin{equation}
\label{eqn:4x4hamiltonian}
\op{H} =
\begin{bmatrix}
0 & v_{3}\op{\pi} & 0 & v\op{\pi}^{\dag} \\
v_{3}\op{\pi}^{\dag} & 0 & v\op{\pi} & 0 \\
0 &  v\op{\pi}^{\dag} & 0 & \xi\gamma_{1} \\
v\op{\pi} & 0 & \xi\gamma_{1} & 0
\end{bmatrix}.
%\op{H} =
%\begin{bmatrix}
%\frac{u}{2} & v_{3}\op{\pi} & 0 & v\op{\pi}^{\dag} \\
%v_{3}\op{\pi}^{\dag} & -\frac{u}{2} & v\op{\pi} & 0 \\
%0 &  v\op{\pi}^{\dag} & -\frac{u}{2} & \xi\gamma_{1} \\
%v\op{\pi} & 0 & \xi\gamma_{1} & \frac{u}{2}
%\end{bmatrix}.
\end{equation}
Above, $\xi=1$ ($\xi=-1$) refers to valley $K$ ($K'$), $v=\frac{a\sqrt{3}\gamma_{0}}{2\hbar}$, $v_{3}=\frac{a\sqrt{3}\gamma_{3}}{2\hbar}$ and $\op{\pi}=p_{x}+ip_{y}$ with the momentum $\vect{p}=(p_{x},p_{y})$ measured from the center of the valley. % We also introduced in Eq.~\eqref{eqn:4x4hamiltonian} the potential energy difference $u$ between the on-site energies in layers 1 and 2 \cite{mccann_landau-level_2006, mccann_asymmetry_2006}.

Using Hamiltonian $\op{H}$ in Eq.~\eqref{eqn:4x4hamiltonian}, the dispersion of the low-energy bands can be written as

\begin{eqnarray}
\epsilon_{\alpha} &=& \alpha\frac{1}{\sqrt{2}}\left\{ \gamma_{1}^{2}+2v^{2}p^{2}+v_{3}^{2}p^{2} - \left[\gamma_{1}^{4}+4\gamma_{1}^{2}v^{2}p^{2}\right.\right.\\ \nonumber
&&-v_{3}^{2}p^{2}(2\gamma_{1}^{2}-4v^{2}p^{2}-v_{3}^{2}p^{2})\\ \nonumber
&&\left.\left.+8\xi\gamma_{1}v_{3}v^{2}p^{3}\cos 3\varphi \right]^{\tfrac{1}{2}} \right\}^{\tfrac{1}{2}}
\end{eqnarray}

where $\alpha=1$ ($\alpha=-1$) denotes the conduction (valence) band and $p=|\vect{p}|$. While for a single-layer of graphene, the in-plane coupling $\gamma_{0}$ leads to the linear Dirac-like dispersion of electrons \cite{wallace_physrev_1947} with the ``effective speed of light'' $v\approx 10^{6}$~m/s \cite{geim_science_2009}, the coupling $\gamma_{1}$ between the two layers turns the linear dispersion into a quasi-parabolic one, giving $\epsilon\approx\pm\frac{p^{2}}{2m}$, at intermediate energies ($\gamma_{1}(\tfrac{v}{v_{3}})^{2}<|\epsilon|\ll\gamma_{1}$) \cite{mccann_landau-level_2006}. The effective mass is given by $m=\frac{\gamma_{1}}{2v^{2}}\approx 0.034m_{e}$, where $m_{e}$ is the free electron mass. The skew interlayer hopping $\gamma_{3}$ introduces trigonal warping of the electronic dispersion, visible in the shape of the constant-energy contour in Fig.~\ref{fig:intro}(c). Most importantly, in unperturbed bilayer graphene it leads at low energies ($|\epsilon|<\tfrac{1}{4}\gamma_{1}(\tfrac{v_{3}}{v})^{2}\equiv\epsilon_{\mathrm{LT}}\approx 1$~meV) to the fragmentation of the singly-connected Fermi line into four separate pieces \cite{mccann_landau-level_2006}, as shown in the right of Fig.~\ref{fig:intro}(c). This is because the band structure at the neutrality point consists of four (anisotropic) mini Dirac cones. Because the energy of the Lifshitz transition $\epsilon_{\mathrm{LT}}$ at which this fragmentation of the Fermi line occurs is very close to the neutrality point, the carrier density required to shift the chemical potential to $\epsilon_{\mathrm{LT}}$ is small, $n_{\mathrm{LT}}\sim 2\times 10^{10}~\mathrm{cm}^{-2}$. In fact, this carrier density is smaller than or comparable to the local charge density fluctuations induced by charged impurities in most of the substrates used for graphene. For that reason, as described in Sec.~\ref{techno}, extremely clean flakes are required to probe the Lifshitz transition in unperturbed bilayer graphene.

The trigonal warping that leads to the Lifshitz transition in bilayer graphene results in a similar spectral feature in the vicinity of the $K$-point of the Brillouin zone of bulk graphite \cite{mcclure1957, orlita_prl_2012}. Because the chemical potential in this case, like for unperturbed bilayer graphene, is close to the Lifshitz transition, signatures of the change of topology of the band structure can be observed for example in cyclotron resonance experiments \cite{orlita_prl_2012}. However, despite the similarities between the band structures of graphite and bilayer graphene, we emphasize that the former lacks the tunability of the latter, both with regards to changing the Fermi level and modifying the Lifshitz transition with external factors.

\subsection{\label{electro_states}Electronic states in the presence of a strong perpendicular magnetic field}

		%%%%% Fig - mag. field theory %%%%%
\begin{figure}[tbp]
\centering
\includegraphics[width=\columnwidth]{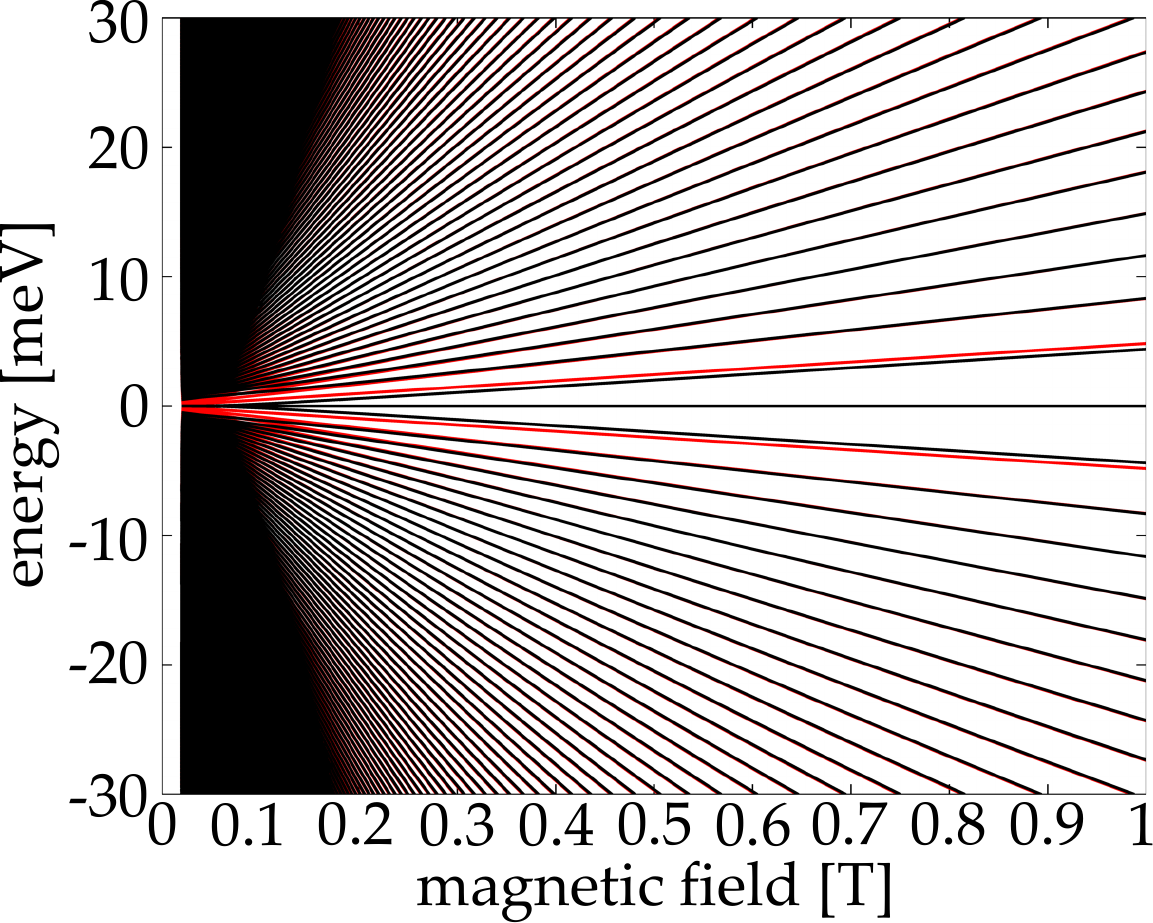}
\caption{Comparison of the Landau level spectra calculated for gapless bilayer graphene without taking into account the $\gamma_{3}$ coupling (solid red lines) and with $\gamma_{3}$ included (solid black lines). For the latter case, we assumed $v_{3}/v=0.1$.}
\label{fig:mag_field_theory}
\end{figure}
		%%%%% End of Fig - mag. field theory %%%%%

In a sufficiently strong magnetic field perpendicular to the graphene layers, the electronic band structure described before undergoes Landau quantization and forms a ladder of discrete energy levels. We describe the theoretical procedure used to calculate the Landau level (LL) structures shown in this paper in Appendix A. In Fig. \ref{fig:mag_field_theory} we show a comparison between the LL spectra calculated using the Hamiltonian in Eq.~\eqref{eqn:4x4hamiltonian}, for $\gamma_{3}=0$ (solid red lines) and $\gamma_{3}\neq 0$ (solid black lines). At high magnetic fields the two spectra are essentially identical, with the energy of a LL $n^{\alpha}$ ($\alpha$ is used to distinguish between the electron and hole levels with the same $n$) described by $\epsilon_{n^{\alpha}}=\alpha\hbar\frac{eB}{m}\sqrt{n(n-1)}$ \cite{mccann_landau-level_2006}. According to this formula, Landau levels $n=0$ and $n=1$ have the same energy, $\epsilon=0$, so that the zero-energy LL, characteristic for graphene, exhibits an extra double orbital degeneracy. This additional degeneracy results in a step of $8\frac{e^{2}}{h}$ in the measured Hall conductance when the Fermi level crosses the neutrality point and suggests that the filling factors $\nu=\pm 4$ display the largest activation gap \cite{mccann_landau-level_2006, novoselov2006}. The higher LLs form an almost equidistant staircase of electronic states and are fourfold degenerate due to the valley and spin degeneracies. In high quality bilayer graphene, strong electron-electron interaction can however lift these degeneracies, as will be discussed later.

As the low-energy band structure contains four mini Dirac cones, at very low magnetic fields the zero-energy Landau level has an additional 16-fold degeneracy (four Dirac cones $\times 2$ valleys $\times 2$ spins). However, already at magnetic fields $B\approx 0.1~\rm{T}$,  the inverse of the magnetic length $\lambda_{B}=\sqrt{\frac{\hbar}{eB}}$ is comparable with the distance in momentum space between the mini Dirac points. This leads to magnetic breakdown \cite{cohen_prl_1961, blount_physrev_1962} as the separate pieces of the Fermi line are no longer resolved individually. As a result, the 16-fold degeneracy of the zero-energy Landau level is lifted at $B \approx 0.1~\rm{T}$ and the high-field structure described above is formed. Due to the presence of the Lifshitz transition, in high quality samples one might expect filling factors $\nu = \pm 8$ to be the most robust at the lowest magnetic fields.

In this theoretical part, we described the role of each interatomic coupling constant on the band structure and revealed the importance of the skew interlayer coupling $\gamma_{3}$, which gives rise to a Lifshitz transition close to the charge neutrality point. We showed that observing this transition is challenging because it takes place at low energies compared to the usual amount of disorder present in bilayer graphene samples. Obtaining high quality samples is therefore one key requirement to experimentally access this phenomenon. In the next section, we will explain how high quality graphene can be obtained in experiments and how its band structure can be influenced by external parameters.

\section{\label{techno}The tunability of bilayer graphene}

Since its discovery, graphene has raised a lot of experimental interest, both from the fundamental point of view (offering, among other things, access to relativistic phenomena in a solid state system) and because of potential electronic applications \cite{geim_science_2009}. However, graphene quickly appeared to be only that ``perfect" in the theory world: graphene, once isolated from graphite flakes, has to be deposited on a substrate and further processed to be experimentally probed. These fabrication steps necessarily introduce disorder. In this part, we focus on explaining how the quality of graphene devices can be improved by finding substrates alternatives to $Si/SiO_{2}$ and how this allows to experimentally access the Lifshitz transition.

\subsection{Substrate-induced disorder in single and bilayer graphene}

In early graphene-based transport experiments, graphene was deposited on $Si/SiO_{2}$ substrates \cite{novoselov2004,novoselov2005,zhang2005}, which offer both good optical contrast to locate graphene and a way to electrostatically tune the Fermi energy using doped silicon as a backgate. However, it was rapidly found that this substrate, as convenient as it is, is a non-negligible source of disorder: the surface is rough, contains impurities, charge traps as well as charged surface states, which all limit the electronic performance of the resulting devices \cite{ando2006,ishigami2007,fratini2008,chen2008}. To circumvent this issue, two main paths have been investigated.

The first one consists in suspending graphene to remove the substrate altogether. Until recently, this yielded the best quality graphene ever achieved \cite{bolotin2008,du2008}. This technique, however, results in fragile devices with complicated architectures if additional electric gates are required. It also limits the range of vertical electric fields that one can apply to graphene before it collapses. At the same time, this approach could potentially be used to induce controlled strain in the suspended graphene membrane \cite{zhang_carbon_2014}. As discussed in the next Section, this leads to interesting consequences for the band structure of bilayer graphene.

In parallel, another option was developed by Dean et al. \cite{dean2010}: using hexagonal boron nitride (h-BN) as a substrate. This material is both a good substrate (low roughness and free of charge traps) and a good dielectric (it is very robust and allows the application of high electric fields). The first process developed by Dean et al., called the ``dry transfer technique'', showed a clear enhancement of the charge carrier mobilities \cite{dean2010}. However, these numbers could still not compete with suspended devices, mostly because of the difficulty to get rid of process-induced residues. But very recently, the same group developed a new technique \cite{wang2013}, simultaneously with other groups \cite{zomer2014}, called the ``pick-up'' technique, allowing for one-dimensional contacts. This new process allows one to create h-BN/Graphene/h-BN stacks that can be contacted at the edge, leaving the graphene free of any polymer residue. This also allows the graphene-on-boron nitride technology to now compete with the quality of suspended devices, with the advantage of being free of strain and to allow the application of large gate voltages through the flake.

Another advantage of the encapsulation of bilayer graphene lies in the fact that one can easily pattern top gates on the surface of the top h-BN: together with the Si-backgate, they allow for the application of a potential asymmetry between the top and bottom layer of the bilayer flake. Such an asymmetry has profound consequences for the band structure of bilayer graphene and, as discussed later in this review, on the Lifshitz transition present in this material.

We should however highlight at this stage that the substrate removal or replacement does not solve all the problems identified by transport experiments: besides the clear improvement of the bulk transport quality, the problems concerning nanostructures remain unchanged \cite{bischoff2012}, as it seems that the performance of these devices is mostly dominated by the edges \cite{bischoff2014}.

In the following, we will explain in more detail the theoretically expected effects on the band structure of strain and of the field-induced asymmetry, and we will show their influence on the Lifshitz transition.

%-------------------------- MARCIN START -------------------------%
\subsection{\label{sec:strain}Influence of strain}

\begin{figure}[tbp]
\centering
\includegraphics[width=\columnwidth]{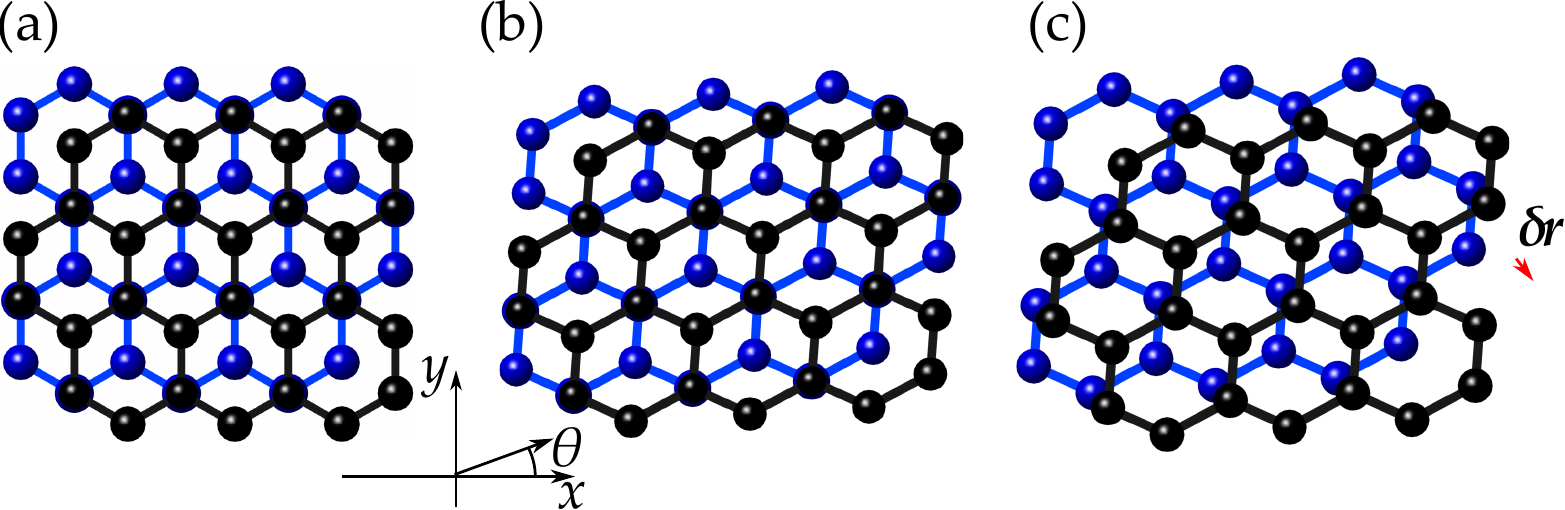}
\caption{Top view of (a) unstrained, (b) strained, (c) strained and displaced graphene layers in bilayer graphene.}
\label{fig:strained_lattice}
\end{figure}

In this section, in particular the influence of (homogeneous) strain on the topology of the low-energy bands and the Lifshitz transition energy $\epsilon_{\mathrm{LT}}$. Strain can potentially be induced in suspended flakes, either intentionally \cite{zhang_carbon_2014} or unintentionally, for example as a side effect of the fabrication process. We neglect here the out-of-plane deformations of the graphene sheets and consider an undeformed bilayer graphene crystal, shown in Fig.~\ref{fig:strained_lattice}(a), which undergoes homogeneous deformation characterized by an angle $\theta$ between the principal axis of strain tensor and the $x$-axis of the coordinate system as depicted in Fig. \ref{fig:strained_lattice}(b), and its eigenvalues, $\delta$ and $\delta'$. Note that after such a deformation, the atomic sites $A2$ and $B1$ remain on top of each other. We further include shear deformations by allowing a shift of the top layer with respect to the bottom one, by $\vect{\delta\!r}=(\delta\!x,\delta\!y)$, as shown in Fig.~\ref{fig:strained_lattice}(c). As a result of homogeneous strain, (1) the Brillouin zone is distorted due to the real space deformation of the lattice, (2) distances between carbon atoms change implying modification of the hopping integrals. The former is a purely geometrical effect leading to a shift of the valleys from their position in the reciprocal space. The latter means that the cancellation between the contributions of the three $\gamma_{0}$ ($\gamma_{3}$) hops for the coupling of plane wave Bloch states on $A1/A2$ and $B1/B2$ sublattices in the (shifted) valleys $K$ and $K'$ of the Brillouin zone is no more exact, leading to the Hamiltonian \cite{mucha-kruczynski_prb_2011, mucha-kruczynski_ssc_2011}

\begin{equation}
\label{eqn:4x4hamiltonian_strain}
\op{H} =
\begin{bmatrix}
0 & v_{3}\op{\pi} + w & 0 & v\op{\pi}^{\dag} \\
v_{3}\op{\pi}^{\dag} + w^{*} & 0 & v\op{\pi} & 0 \\
0 &  v\op{\pi}^{\dag} & 0 & \xi\gamma_{1} \\
v\op{\pi} & 0 & \xi\gamma_{1} & 0
\end{bmatrix},
\end{equation}
 where
\begin{align}
\label{eqn:w}
w = \frac{3}{4}(\eta_{3}-\eta_{0})\gamma_{3}e^{-i2\theta}(\delta-\delta ')-\frac{3}{2} \gamma_{3}\eta_{3}e^{i\varphi}\frac{\delta r}{r_{AB}},\,\,\,\,\tan\varphi=\frac{\delta\!y}{\delta\!x}. \nonumber
\end{align}
Above, $r_{\mathrm{AB}}$ is the equilibrium carbon-carbon distance and $\eta_{0}=\frac{r_{\mathrm{AB}}}{\gamma_{0}}\frac{\partial\gamma_{0}}{\partial r_{\mathrm{AB}}}$ and $\eta_{3}=\frac{r_{AB}}{\gamma_{3}}\frac{\partial\gamma_{3}}{\partial r_{AB}}$ quantify the change of the intralayer $A$-$B$ and interlayer $A1$-$B2$ hoppings, respectively, upon the change of the distance between the coupled carbon atoms. Note that $w$ is in general a complex number with the phase specified by the directions of the principal axis of strain and the shear deformation. In the case of the former, rotation by $\pi$ leaves $w$ unchanged (it does not matter by which of the opposite edges the material is stretched). Also, for the same eigenvalues of the strain tensor and ignoring shear, applying strain along the zigzag or armchair directions is equivalent to switching the sign of $w$. For shear deformation, note that to simplify notation the angle $\varphi$ is defined as the one between the negative direction of the $y$ axis and the direction of shear.

Based on the analysis of Raman spectra on monolayer graphene, it has been estimated \cite{mucha-kruczynski_prb_2011} that 1\% of strain can lead to $w\sim 6$~meV. In the same work it has also been established that electron-electron interaction would further enhance the strain-induced effects. This can be compared to the Lifshitz transition energy $\epsilon_{\mathrm{LT}}\approx1$~meV in the unperturbed bilayer graphene.

		%%%%% Fig - strained lattice %%%%%
		%%%%% End of Fig - strained lattice %%%%%

\begin{figure*}[tbp]
\includegraphics[width=\textwidth]{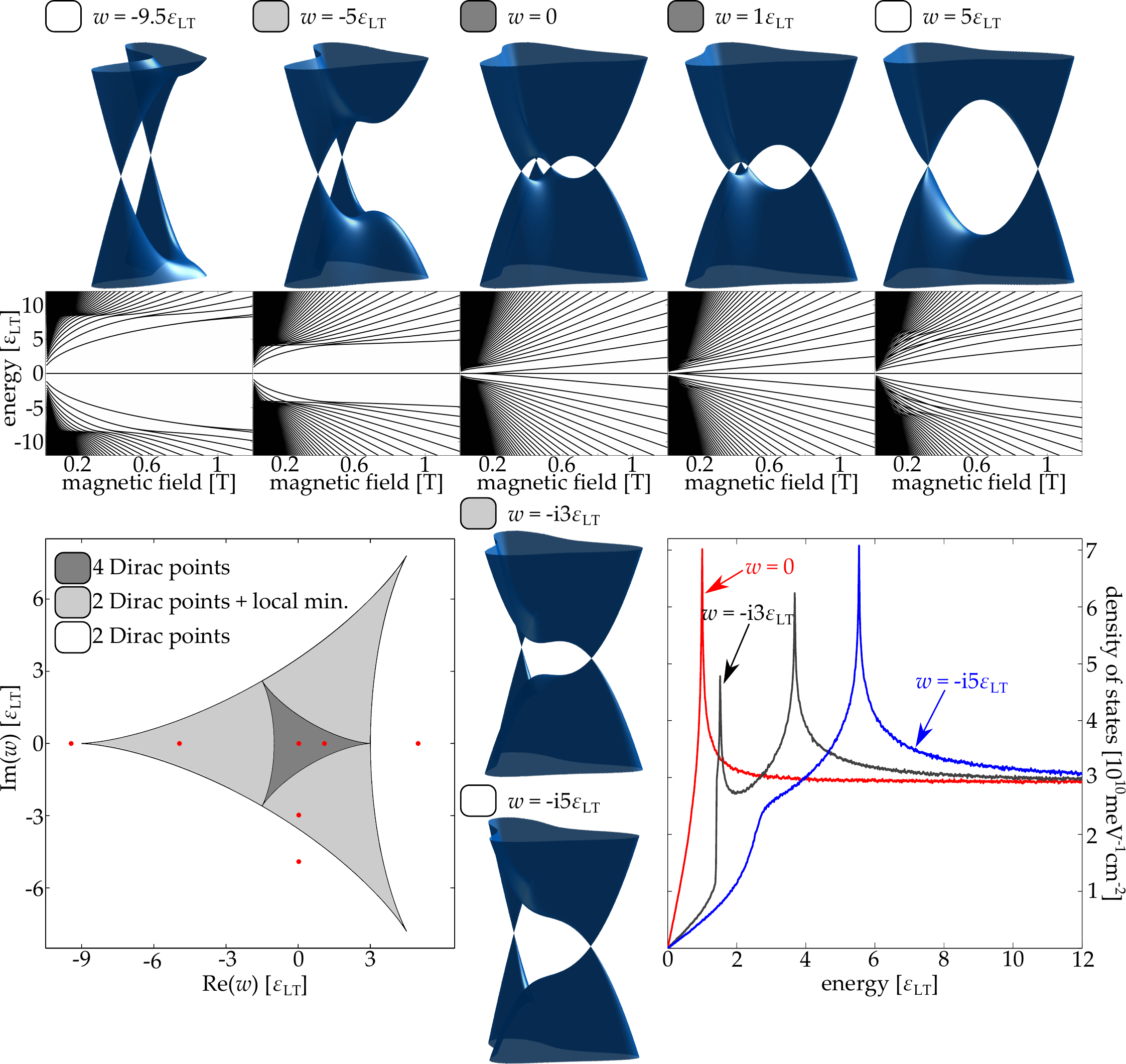}
\caption{The low-energy band structures for strained bilayer graphene. For the band structures shown in the horizontal row, the corresponding Landau level fans are shown underneath. Densities of states calculated for the electronic spectra displayed in the vertical column in the center are shown in the bottom right. Each electronic spectrum is assigned a label colored according to the characteristic topology of the spectrum. The extent of the parametric regimes of complex $w$, see Eq.~\eqref{eqn:4x4hamiltonian_strain}, distinguishing between three characteristic topologies of the strained bilayer graphene spectrum is displayed using those colors in the bottom left. In this graph, red dots show the points in the parametric space corresponding to the spectra presented in this figure.}
\label{fig:strain_bands}
\end{figure*}

The resulting low-energy electronic dispersions of strained bilayer graphene in the vicinity of the valley $K$ are displayed for representative values of $(\Re w,\Im w)$ in Fig. \ref{fig:strain_bands}. The dispersion for the case of the unperturbed bilayer graphene, that is for $w=0$, is shown in the center of the top row in Fig.~\ref{fig:strain_bands}. As mentioned before, it is quasi-parabolic at energies $\epsilon\gg\epsilon_{\mathrm{LT}}=\frac{mv_{3}^{2}}{2}$, with $v_{3}$ responsible for the trigonal warping of isoenergetic lines. At $\epsilon=\epsilon_{\mathrm{LT}}$, the isoenergetic line undergoes a Lifshitz transition and splits into four Dirac cones \cite{mccann_landau-level_2006}. The remaining dispersions in the top row depict the band structure in the vicinity of the neutrality point for $\Re w\neq 0$, $\Im w=0$. As the value of $\Re w$ is increased (right side of the top row in Fig.~\ref{fig:strain_bands}), the two side cones positioned off the $\px$ axis and the central cone move closer, as shown in the graph for $w=\epsilon_{\mathrm{LT}}$. Those three cones collide for $w=3\epsilon_{\mathrm{LT}}$, and for $w>3\epsilon_{\mathrm{LT}}$ only two Dirac cones remain, as shown in Fig.~\ref{fig:strain_bands} for $w=5\epsilon_{\mathrm{LT}}$. If $\Re w<0$ is negative (left side of the top row in Fig.~\ref{fig:strain_bands}), the central cone and the side cone positioned on the $x$ axis approach each other and collide for $w=-\epsilon_{\mathrm{LT}}$, creating a local, quasi-parabolic minimum. This minimum persists for some range of the strain, although it lifts off the $\epsilon=0$ plane, as shown in the graph for $w=-5\epsilon_{\mathrm{LT}}$. Eventually, for $\Re w=-9\epsilon_{\mathrm{LT}}$, the minimum merges with a saddle point and with further decrease of $\Re w$, again only two Dirac cones remain in the spectrum, as shown in the spectrum for $w=-9.5\epsilon_{\mathrm{LT}}$. A contrasting situation of $\Re w=0$, $\Im w\neq 0$, is presented in the graphs in the column in the middle of Fig.~\ref{fig:strain_bands} (due to symmetry, dispersions for opposite values of $\Im w$ are mirror reflections of each other with respect to the $\py=0$ plane and hence only situation of $\Im w<0$ is described here). Again, starting from the unperturbed system and increasing the magnitude of $\Im w$ leads to two of the cones moving closer to each other and colliding with a creation of a local, quasi-parabolic minimum which lifts off the $\epsilon=0$ plane (see the dispersion for $w=-i3\epsilon_{\mathrm{LT}}$) and eventually disappears, leaving only two Dirac cones (graph for $w=-i5\epsilon_{\mathrm{LT}}$). Three qualitatively different regimes in the $(\Re w,\Im w)$ space can be identified: (1) the spectrum contains four Dirac cones, (2) the spectrum contains two Dirac cones and a local minimum, created by a collision of two of the cones and which in general does not touch the $\epsilon=0$ plane, and (3) the spectrum contains only two Dirac cones. The extent of those regimes is shown in the bottom left of Fig.~\ref{fig:strain_bands}, together with the points in this space corresponding to spectra discussed above and shown in Fig.~\ref{fig:strain_bands}. Note that symmetry of the lattice requires that, for no shear, rotating the principal axes of strain by $\pi/3$ leads to the same deformation. In the case of pure interlayer shear, choosing directions related by $2\pi/3$ rotation also results in the same physical situation. Both are reflected by the $2\pi/3$ symmetry of the graph in the bottom left in Fig.~\ref{fig:strain_bands} (as the angle $\theta$ of applied strain appears doubled in the definition of $w$).

		%%%%% Fig - strain bands %%%%%

		%%%%% End of Fig - strain bands %%%%%

The strain-induced deformation of the low-energy electronic dispersion gives rise to distinctive features in the low-energy density of states (DoS) in each of the strain regimes. In particular, existence of a local parabolic minimum results in a sharp step in the DoS, while saddle points lead to van Hove singularities. Note that for $w\neq 0$, the symmetry between the side cones is broken and the saddle points are no longer all found at the same energy $\epsilon_{\mathrm{LT}}$, leading to more than one van Hove singularity and more than one Lifshitz transition. Densities of states representative of all the regimes are shown in the bottom right of Fig.~\ref{fig:strain_bands} for $w=0$,  $w=-i3\epsilon_{\mathrm{LT}}$ and $w=-i5\epsilon_{\mathrm{LT}}$, that is those of the points marked in red in the diagram in the bottom left of Fig.~\ref{fig:strain_bands} that lie on the $\Re w=0$ line. Indeed, for $w=0$ only one van Hove singularity is present, whereas two can be seen for $w=-i3\epsilon_{\mathrm{LT}}$. A sharp step, corresponding to the contribution of a quasi-parabolic part of the band structure to the density of states, is also visible for the latter case. Finally, for $w=-i5\epsilon_{\mathrm{LT}}$, the sharp step is washed out (as the local minimum merged with the saddle point) and only one van Hove singularity remains.

The transformation of electron dispersion by homogeneous strain leads to the modification of the bilayer graphene LL spectrum. The examples of numerically calculated LLs are shown for the electronic dispersion in the top row of Fig.~\ref{fig:strain_bands} for low magnetic fields, $B\leq 1$T. Both for small and large strain, the high-magnetic-field end of the LL fan plot can be described neglecting the trigonal warping, as mentioned in Section~\ref{electro_states}. At low fields, such that $\hbar\omega_{c}(B)<mv_{3}^{2}$, the largest gap in the LL spectrum is between the $\epsilon=0$ and next excited LL, suggesting the persistence of filling factor $\nu=\pm 8$ in the quantum Hall effect (QHE). After strain causes the annihilation of two out of four Dirac points, the $\epsilon=0$ level becomes eightfold degenerate, and, for strain $|w|\gg\epsilon_{\mathrm{LT}}$, only filling factors $\nu=+4$ and $\nu=-4$ persist in the low-field QHE in bilayer graphene: the largest energy gap in the LL spectra is between the eightfold degenerate level at $\epsilon=0$ and next excited level, whereas the rest of the spectrum is quite dense. This eightfold degeneracy is topologically protected and it also appears in a rotationally twisted two-layer stack \cite{de_gail_prb_2011}. Importantly, for intermediate strains such that the low-energy electronic dispersion displays a quasi-parabolic local minimum, the LL structure contains a Landau level origins of which can be traced to electronic states in that minimum - its dispersion as a function of the magnetic field $B$ is linear and its energy with vanishing $B$ does not go to zero but rather follows the position of said minimum.
%-------------------------- MARCIN END -------------------------%

\subsection{Influence of the interlayer asymmetry $u$}
As mentioned earlier, the encapsulation of bilayer graphene enables the use of top gates. This allows applying a high field-induced asymmetry between the two layers which strongly influences the band structure..
\subsubsection{\label{mexico}Inducing an asymmetry $u$ with $\gamma_{3}=0$ : the Mexican hat}

One of the most interesting things about bilayer graphene as a material is the fact that, since it possesses two layers, one can try to address each layer independently. To do so, one can tune the bottom layer to a potential $u/2$ and the top layer to a potential $-u/2$. This way, an asymmetry of amplitude $u$ is induced between the two layers. The resulting Hamiltonian is now given by

\begin{equation}
\label{hamilt-with-u}
\op{H} =
\begin{bmatrix}
\frac{\xi}{2} u & 0 & 0 & v\op{\pi}^{\dag} \\
0 & -\frac{\xi}{2} & v\op{\pi} & 0 \\
0 &  v\op{\pi}^{\dag} & -\frac{\xi}{2} & \xi\gamma_{1} \\
v\op{\pi} & 0 & \xi\gamma_{1} & \frac{\xi}{2} u
\end{bmatrix}.
\end{equation}

\begin{figure}
\includegraphics[width=\columnwidth]{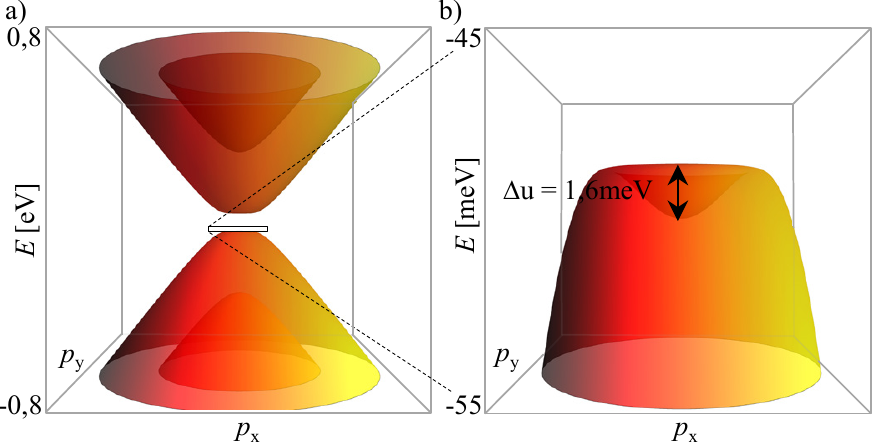}
\caption{(a) Effect of the interlayer asymmetry on the low energy band structure of bilayer graphene in the proximity of the $K$-point, calculated taking into account only the in-plane coupling $\gamma_{0}$ and the interlayer coupling between dimers $\gamma_{1}$ ($\gamma_{3}=0$): a band gap is opened (here $u$ is set to $100~\rm{meV}$). (b) Zoom at the top of the valence band: the so-called ``Mexican hat'' is visible. For $u = 100~\rm{meV}$, the amplitude of the hat is $\Delta u = 1.6~\rm{meV}$.}
\label{fig1}
\end{figure}

\begin{figure}[h]
\includegraphics[width=\columnwidth]{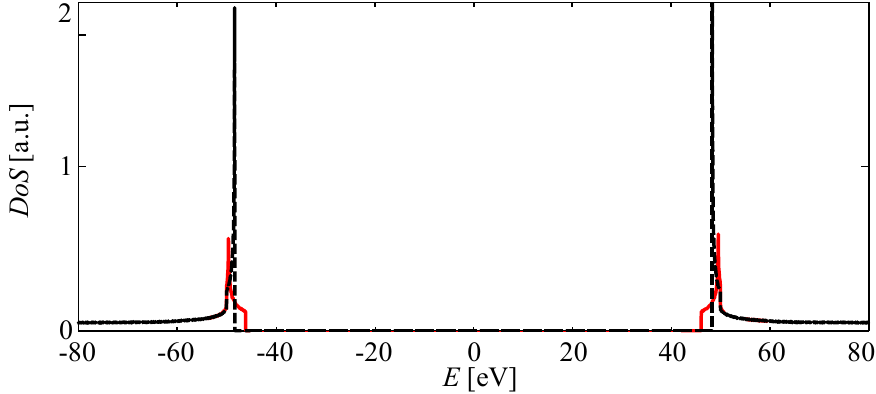}
\caption{Density of state calculated for the case with trigonal warping included (red curve) and without ($\gamma_{3} = 0$, black dashed curve). The height of the peaks in the DoS was limited for visual purposes to enable better comparison between other features in the graph.}
\label{figdos}
\end{figure}

To induce this asymmetry, chemical doping can be used \cite{castro_biased_2007} or external gates \cite{Oostinga2008} can be patterned. The relation between $u$ and the applied voltages has been investigated in Ref.~\cite{mccann_asymmetry_2006,min2007}. This induced asymmetry opens a band gap, as shown in Fig.~\ref{fig1}(a) \cite{mccann_landau-level_2006,mccann_asymmetry_2006,mccann2007,mucha-kruczynski_influence_2009}. The gap gives rise to a characteristic shape of the conduction/valence band edge, a so called ``Mexican hat''. This is shown in Fig.~\ref{fig1}(b). For an induced asymmetry $u = 100~\rm{meV}$, the amplitude of the central dip is $\Delta u = 1.6~\rm{meV}$. The corresponding density of states is shown in Fig.~\ref{figdos} (black dashed curve).

This asymmetry also has consequences for the magnetic field behavior, as it lifts the valley degeneracy of the Landau levels \cite{mccann_landau-level_2006}. This is because the Landau level wave function is not distributed equally among the layers, with the most striking example being that of the zero-energy Landau level. For this level, electronic states in the valley $K$ are formed by orbitals that sit mostly (and exclusively in the case of gapless bilayer) on the $A1$--sites, while in the valley $K'$, these states occupy dominantly the $B2$ sublattice. Thus, since for nonzero interlayer asymmetry these sites are set to different potentials, electronic states in $K$ are no longer degenerate with states in $K'$. Such lifting of the valley degeneracy occurs for all Landau levels but the magnitude of the energy splitting decreases with increasing Landau level index \cite{mccann_landau-level_2006}. In experiments, even though the lifting of all the degeneracies has been observed for the lowest LL, temperature and energy broadening usually hinder the splitting for higher LLs.

\subsubsection{\label{theory_asymmetry_and_trig}Influence of $\gamma_{3}$ in a gapped bilayer graphene system}

\begin{figure*}[tbp]
\includegraphics[width=\textwidth]{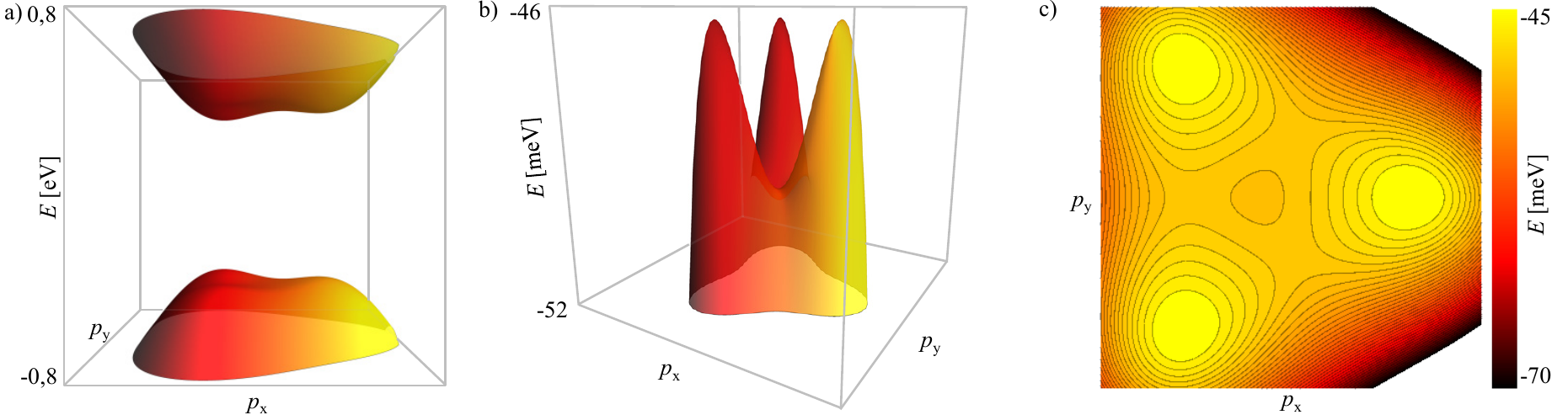}
\caption{(a) Band structure obtained while applying an interlayer asymmetry $u = 100~\rm{meV}$ on a bilayer graphene and taking into account $\gamma_{3}$: close to the gap, the trigonal distortion is still visible. (b) Zooming at the top of the valence band, one can clearly observe that there are three outer maxima and a minimum in the center. The energy between them is around $3~\rm{meV}$ for $u = 100~\rm{meV}$. (c) Taking constant energy cuts in (b), one can observe that, again, as a function of energy, the Fermi contour gets broken, this time into three pockets: this illustrates the Lifshitz transition in a gapped bilayer graphene system.}
\label{fig3}
\end{figure*}

As mentioned previously, accessing experimentally the Lifshitz transition in bilayer graphene is challenging. However, by using the possibility of applying a vertical electric field, one can realize such a situation experimentally.

Considering the following electron-hole symmetric Hamiltonian,

\begin{equation}
\label{hamilt-with-u-full}
\op{H} =
\begin{bmatrix}
\frac{\xi}{2} u & v_{3}\op{\pi} & 0 & v\op{\pi}^{\dag} \\
v_{3}\op{\pi}^{\dag} & -\frac{\xi}{2} & v\op{\pi} & 0 \\
0 &  v\op{\pi}^{\dag} & -\frac{\xi}{2} & \xi\gamma_{1} \\
v\op{\pi} & 0 & \xi\gamma_{1} & \frac{\xi}{2} u
\end{bmatrix},
\end{equation}

one can now set the interlayer asymmetry to $u = 100~\rm{meV}$ and observe the consequences on the band structure. As shown in Fig.~\ref{fig3}(a), the band gap is open in a similar way as in the case where no skew interlayer hopping was included. However, already on this scale of $\pm0.8~\rm{meV}$, one can see that the trigonal perturbation is still present. Zooming at the top of the valence band, as done in Fig.~\ref{fig3}(b), we see that the band gap opening has a strong effect on the previous four Dirac cones: only three peaks are left and they do not exhibit a linear dispersion anymore. The central peak has been pushed down by the $\gamma_{1}$ parameter, by the same process as for the ``Mexican hat''. The energy contours however show the same kind of physics as a function of energy. Fig.~\ref{fig3}(c) shows these contours as a function of energy for the valence band: again, at high negative energies, the band structure exhibits one unique and continuous energy contour. While rising the Fermi level towards the gap, the contour breaks up into three pockets, revealing the presence of a Lifshitz transition \cite{lifshitz}. This time, however, the energy scale is bigger: the band gap stretches the three outer peaks and the difference between the three maxima at the top of the valence band and the minimum in the center is now around $\epsilon_{\mathrm{LT}} = 3~\rm{meV}$, as illustrated in Fig.~\ref{fig3}(b). One can therefore conclude from this estimate that the experimental observation of the Lifshitz transition should be facilitated by the layer-asymmetry induced band gap and that, the wider the gap, the more enhanced the visibility of this effect. In Fig.~\ref{figdos}, we additionally show the density of states of such a system. This is highlighted with the red solid curve, which is to be compared with the black dashed curve corresponding to the case where no trigonal warping was included ($v_{3} = 0$).

This new topology influences the quantum Hall effect as well. Instead of the extra fourfold symmetry arising from the four Dirac cones, the triplet at the top of the valence band gives rise to a threefold degeneracy \cite{varlet_anomalous_2014}. Because of the induced band gap, the valley degeneracy is lifted and therefore the plateaus that would survive down to the lowest magnetic fields would be $\nu = \pm6$ and, because of exchange interaction, $\nu = \pm3$. The full LL spectrum will be shown and discussed in more details in Section~\ref{results}.

In this section, we pointed out how the low energy dispersion of bilayer graphene can be manipulated by external parameters. Effects on the Lifshitz transition are strong and can in principle be experimentally observed by studying the degeneracies of the Landau level spectrum. In the next Section, we will illustrate how the proposed effects are seen in experiments by focusing on the effect of the interlayer asymmetry.

\section{\label{asymmetry}Probing the Lifshitz transition in gapped bilayer graphene}

As mentioned in the previous part, a way to tune and observe the Lifshitz transition in experiments is to apply an interlayer asymmetry to the bilayer graphene flake. To do so, one needs to apply a vertical displacement field $D$ using a combination of top- and backgate. Encapsulating the bilayer graphene is therefore required. As the results which will be presented in this section have been obtained on a device fabricated from the dry transfer technique \cite{dean2010}, we will first shortly describe this fabrication process, but the reader should keep in mind that recent improvements may lead to even better performances \cite{wang2013,zomer2014}. The electrostatics of the device will be explained as well as the gap properties. In a second part, we will reveal how, using the quantum Hall effect as a tool, one can identify the presence of the Lifshitz transition via the study of the different LL degeneracies.

\subsection{Inducing a gap in bilayer graphene}
\begin{figure}
\centering
\includegraphics[width=\columnwidth]{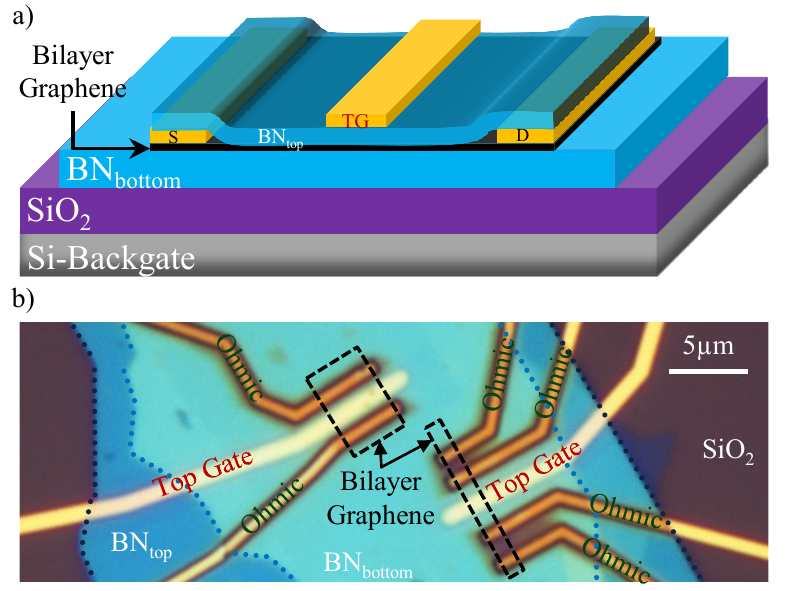}
\caption{(a) Sketch of the device. A contacted bilayer graphene flake is sandwiched between two h-BN flakes. The whole device can be tuned using the $Si$ backgate and the central region can also be independently controlled using the central top-gate. Combining the actions of the top- and backgate, a band gap is open in the central region. (b) Optical Microscope image of the real device.}
\label{fig4}
\end{figure}

The device fabrication starts with the exfoliation of h-BN on $Si/SiO_{2}$. Once a good candidate flake is identified and its cleanliness is confirmed by Atomic Force Microscopy (AFM), a bilayer graphene flake can be transferred on top. The transfer is done, as mentioned earlier, using the dry transfer technique: graphene is exfoliated on top of a $Si$-chip covered with a double-layer resist (in our case PVA + PMMA \footnote{PVA: Polyvinyl alcohol; PMMA: Poly(methyl methacrylate)}), the first layer being water-soluble while the second one is a standard polymer; once the bilayer nature of the flake is confirmed by Raman spectroscopy and its cleanliness checked with AFM, the chip can be put in water: the sacrificial layer is dissolved and we are left with the graphene on PMMA floating at the surface. This floating polymer piece can then be ``fished'', heated up and brought in contact with the h-BN using a micromanipulator. The flake is then etched and contacted using standard electron beam lithography. However, at the end of this step, the graphene surface is usually covered with resist residues. Different cleaning techniques have been established: thermal annealing \cite{dean2010,lin2012}, current annealing \cite{moser2007} or mechanical cleaning \cite{goossens2012}. The presented device has been cleaned using the last technique. Once no resist residue can be identified on the surface, the device is encapsulated: a second h-BN flake is transferred on top in the same way as the graphene, and a top-gate patterned with electron beam lithography in the middle. The resulting device is sketched in Fig.~\ref{fig4}(a) and shown in Fig.~\ref{fig4}(b).

\begin{figure}
\centering
\includegraphics[width=\columnwidth]{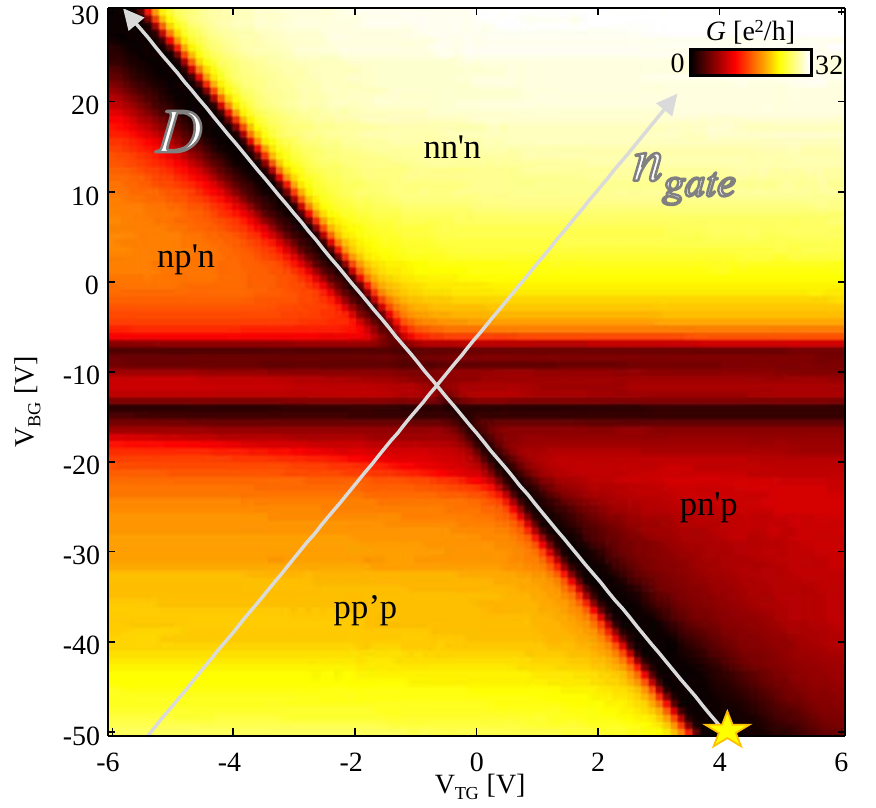}
\caption{Conductance map measured as a function of the voltages applied to the backgate, $V_\mathrm{BG}$, and the one applied to the top-gate, $V_\mathrm{BG}$. The star shows the displacement field value $D = -0.9~\rm{V/nm}$, which is commented on in the main text.}
\label{fig5}
\end{figure}

Such a device defines three regions in series: the two outer ones, close to the Ohmic contacts, are bilayer graphene regions which can only be tuned using the doped silicon backgate, while the central area, under the top-gate, can be tuned combining the top- and backgate voltages (the carrier types -- $n$ or $p$ -- belonging to this area are labeled with a prime in Fig.~\ref{fig5}). The use of these two gates also allows the application of an interlayer asymmetry $u$ and therefore allows opening a band gap in this region.

In Fig.~\ref{fig5}, the electrical characteristics of such a device are shown: the conductance is measured as a function of top- and backgate voltages. Two horizontal lines of low conductance are seen: they represent the charge neutrality in the two untop-gated regions: at the top of these lines the two external regions of the device are $n$-doped and while below, they are $p$-doped. Along the diagonal, another low conductance line is seen: this line, which defines the displacement field axis $D$ (e.g. the line along which the asymmetry between top- and bottom-layer is increased), illustrates the situation where the Fermi energy lies in the gap of the central region. At the left of the $D$-axis, this area is $p$-doped while at the right it is $n$-doped. Along this axis however, the band gap size is increased, from no gap at the crossing of the two displayed axes, $D$ and $n_{gate}$, to a gap size of $u \approx 80~\rm{meV}$ at $D = -0.9~\rm{V/nm}$ (intersection of the $D$-axis and $V_\mathrm{BG} = -50~\rm{V}$, highlighted with a star in Fig.~\ref{fig5}).

The estimation of the gap size is done using a self-consistent procedure, as presented in Refs.~\cite{mccann_asymmetry_2006,mucha-kruczynski_influence_2009}. This allows to take into account the charge redistribution between the layers as a function of changing gap size, which leads to a different screening of the applied $D$-field. The good agreement between theoretical prediction and the induced gap size in the fabricated devices was confirmed in the past by ARPES experiments \cite{ohta2006}. From transport experiments, the gap size should nevertheless be accessible by performing temperature dependent measurements (transport through the gap should obey an Arrhenius law). However, as demonstrated in \cite{Oostinga2008,Russo2009,Weitz2010,Thiti2010}, transport through the gap is actually dominated by hopping processes, which obscure the true gap size. There is therefore no experimental way to directly confirm the estimated gap size in the device. We can nevertheless confirm that our estimation of $u = 80~\rm{meV}$ is not too far from reality, as confirmed both by the quantum Hall effect measurements which will be shown in the next section and by other experiments carried out on the same device, namely the observation and characterization of Fabry-P\'{e}rot interference \cite{varlet_fabry_2014}, the periodicity of which depends on the gap size.

The strength of such an encapsulated geometry lies in the following arrangement: the h-BN material is so robust against electric fields that it is experimentally possible to open such large gaps, by applying such large voltages. This is for example not possible with suspended devices, which would collapse due to electrostatic forces. In this case, the flake remains in a good state and we are able to investigate this high displacement field regime. Every requirement is met to probe the Lifshitz transition.

\subsection{\label{results}Probing the Lifshitz transition using the quantum Hall effect}

As mentioned in Section~\ref{theory_asymmetry_and_trig}, the influence of $\gamma_{3}$ on the band structure leads to a situation amenable for the observation of a Lifshitz transition: from one unique and continuous Fermi contour at positive or negative energy to a broken contour with three separate pockets close to the edges of the bands. This leads to consequences for the quantization observed in the quantum Hall regime: at low magnetic-field a sixfold degeneracy should be observed (three without spin degeneracy). Thus a way to confirm the presence of the Lifshitz transition is to perform quantum Hall measurements on the previously presented device. The results discussed in this section can be found in more detail in reference \cite{varlet_anomalous_2014}.

\begin{figure}
\centering
\includegraphics[width=\columnwidth]{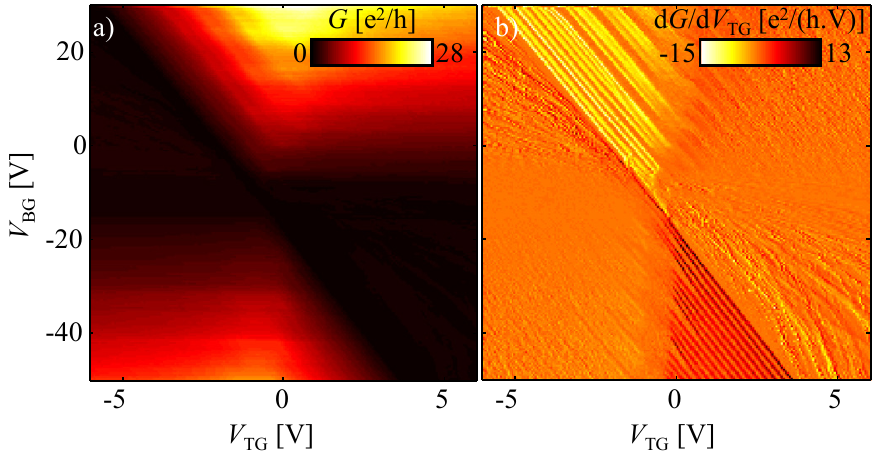}
\caption{(a) Conductance map at $6~\rm{T}$ measured as a function of the voltages applied to the backgate, $V_\mathrm{BG}$, and the one applied to the top-gate, $V_\mathrm{TG}$. (b) Measured normalized transconductance map: a number of lines are revealed at the transition between quantum Hall plateaus.}
\label{fig6}
\end{figure}

Fig.~\ref{fig6}(a) shows the conductance map taken at $6~\rm{T}$. This map is not qualitatively different from the map shown in Fig.~\ref{fig5}. However the conductance does get quantized and this becomes more visible when recording the normalized transconductance signal $dG/dV_{\mathrm{TG}}$: two triangular regions are seen, exhibiting a series of strong lines running parallel to the displacement field axis. These lines delimit a regime in which the density in the two outer regions is bigger than the density in the central one, e.g. more edge modes are present in these two regions ($\nu > \nu'$). Hence, the measured conductance, which reflects the amount of edge modes which can travel through the whole device, will be given by $G = e^{2}\nu'/h$ \cite{williams_quantum_2007,ozyilmaz_electronic_2007,jing_quantum_2010,amet_gate_2013}: in this regime, the measured conductance, and therefore transconductance, reflects the quantum Hall effect in the dual-gated region. One can then interpret each of these lines as the transition between plateaus of quantized conductance belonging to the gapped central area.

\begin{figure}
\centering
\includegraphics[width=\columnwidth]{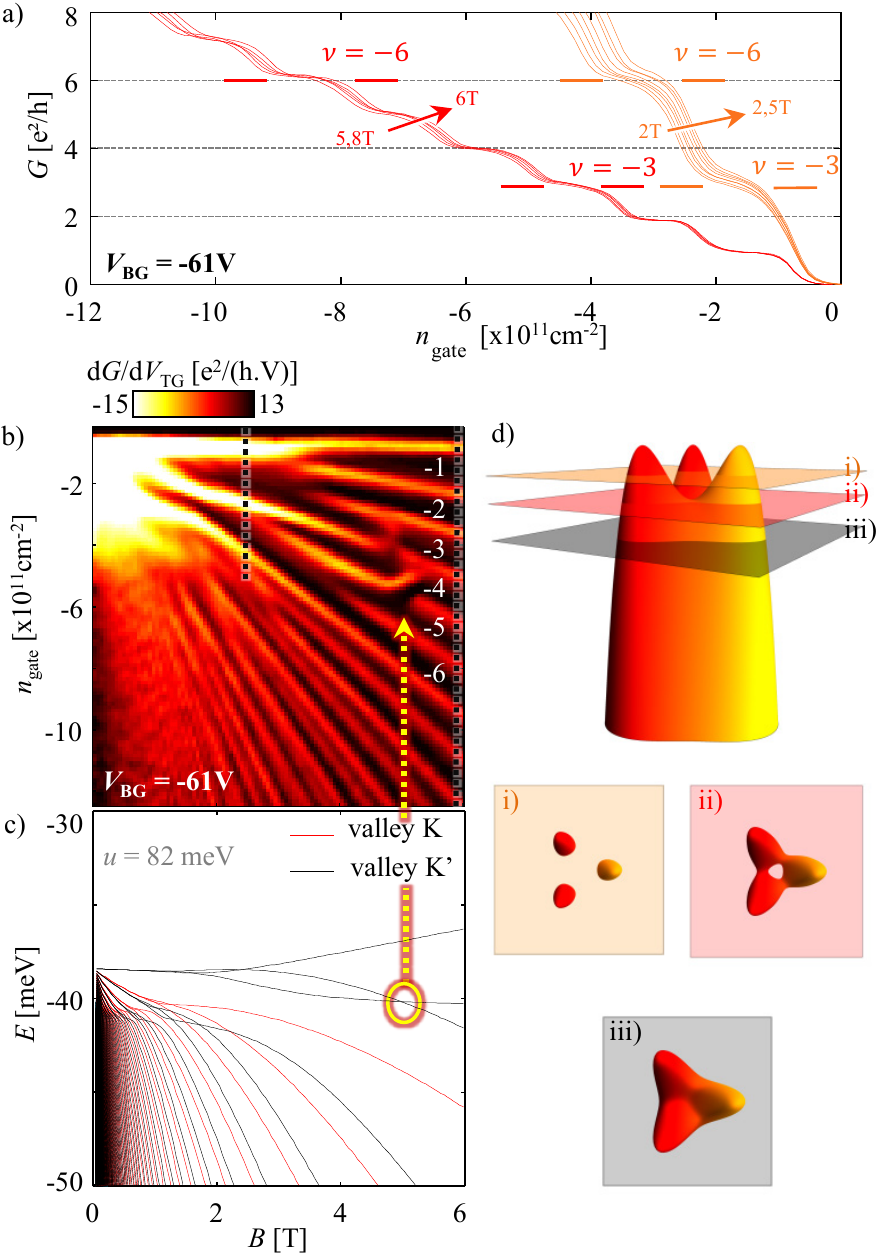}
\caption{(a) Conductance cuts taken along $V_\mathrm{BG} = -61~\rm{V}$, measured sweeping $V_\mathrm{TG} \sim n_\mathrm{gate}$, at two different magnetic field ranges: the measurements reveal a lifting of all the degeneracies at high magnetic fields and a threefold degeneracy at lower fields. (b) Measured Landau level spectrum, showing the evolution of the red conductance cuts from (a) as a function of magnetic field. (c) Corresponding calculated Landau level spectrum, obtained for a gapped trigonally-warped bilayer graphene system, with an interlayer asymmetry set to $u = 82~\rm{meV}$. (d) Valence band of the bilayer graphene, close to the top of the valence. Some representative energy cuts are taken from it to highlight the possible orbital degeneracies of the system.}
\label{fig7}
\end{figure}

To make this more obvious, we investigate a conductance cut at high displacement fields ($V_\mathrm{BG} = -61~\rm{V}$). Fig.~\ref{fig7}(a) shows in red the corresponding conductance measured close to $6~\rm{T}$: it exhibits plateaus at multiples of $e^{2}/h$, indicating that all the degeneracies are lifted at high magnetic fields. This reveals the presence of strong electron-electron interaction and therefore the good quality of the sample. However, a similar behavior is not observed on the electron side, where the LLs remain 4-fold degenerate. In the following, we will therefore focus on the hole side of the dual-gated region.

In Fig.~\ref{fig7}(b), the Landau level spectrum, measured between $0$ and $6~\rm{T}$ at $V_\mathrm{BG} = -61~\rm{V}$, is shown, allowing one to follow the evolution of our broken-symmetry states as a function of magnetic field. This time, the normalized transconductance is displayed and the plateaus are therefore represented by the zero-transconductance regions. Going from high magnetic fields to lower values, one notices a first unexpected feature: following filling factors $\nu' = -4$ and $\nu' = -5$, it appears that, close to $5~\rm{T}$, these two quantum Hall states merge. Going to the lowest magnetic fields ($B \approx 2.5~\rm{T}$) where plateaus are still present, two strong plateaus are seen. Taking cuts around this magnetic field value (yellow dashed line) and displaying the conductance, as done in Fig.~\ref{fig7}(a) (orange cuts), reveals that these two plateaus are the quantum Hall states $\nu' = -3$ and $\nu' = -6$. This threefold symmetry indicates the presence of three spin-split orbits at the same energy, which could be related, as mentioned earlier, to the presence of the Lifshitz transition in the gapped bilayer graphene system.

To get a better insight into these observations, Fig.~\ref{fig7}(c) shows the calculated Landau level spectrum corresponding to bilayer graphene, obtained by taking into account the presence of the gap (calculated self-consistently according to the $D$- and $B$-field values) and the influence of the skew interlayer hopping $\gamma_{3}$ (see \ref{AppB} for more details). This calculation assumes spin-degeneracy, meaning that each branch in the spectrum would count twice if compared to the above measurement. Starting from low magnetic fields, this spectrum exhibits a threefold orbital degeneracy corresponding to the gap at $\nu' = -6$ in the experiment. As mentioned earlier, this is due to the presence of the triplet at the top of the valence band, which gives rise, in magnetic field, to three equivalent orbits, separately confined within the three pockets defined by this triplet (see Fig.~\ref{fig7}(d)i). We tentatively ascribe the observed gap at $\nu' = -3$ to lifted spin degeneracy due to exchange interaction. While increasing the magnetic field, the spin-degenerate orbits grow bigger and start to mix in momentum space, leading to a splitting of the triplet, until this degeneracy is finally fully lifted (already at about $4.6~\rm{T}$ in the experiment). At even higher fields ($B = 6~\rm{T}$) only one unique orbit is left (to be compared with Fig.~\ref{fig7}(d)iii). In-between these two fields, the Lifshitz transition takes place: in the band structure, as shown in Fig.~\ref{fig7}(d)ii, two contours coexist: the Fermi sea has, in its center, an electron-like island. This gives rise to two separated counter-propagating orbits which do not mix and hence lead to an occasional orbital degeneracy (point encircled in Fig.~\ref{fig7}(c): in the calculated spectrum, the gap of $\nu' = -4$ closes and correspondingly the quantized plateau in the experiment disappears (dotted arrow). The position of this crossing corresponds fairly well to the position of the measured one (see yellow arrow), indicating that the value of the gap in the system must be quite close to the estimated one. The investigation of the Landau level degeneracies is therefore an appropriate tool to locate the presence of the Lifshitz transition in bilayer graphene. Beyond that, it is observed that odd filling factors $\nu' = -3$ and $\nu' = -5$ remain quantized at the expected values as the Landau levels cross. Also, $\nu' = -3$ vanishes at magnetic fields slightly below the field of the Lifshitz transition. This behavior of the odd filling factors may be due to interaction effects and remains to be investigated further at a theoretical level.

\section{Conclusion}
Bilayer graphene is a very tunable material. Small features, such as the Lifshitz transition, can be largely influenced by external parameters such as strain, displacement fields or even magnetic fields. Bilayer graphene represents therefore a unique system in which the topology of the band structure can be externally influenced and chosen.

\section*{Aknowledgements}
We acknowledge financial support from the Marie Curie ITNs $S^{3}NANO$, QNET, the Swiss National Science Foundation via NCCR Quantum Science and Technology, the ERC Synergy Grant 'Hetero$2$D', the EPSRC Grant EP/$L013010$/$1$ and the European Graphene Flagship Project.

\bibliography{mybibfile}

\begin{thebibliography}{63}%
\makeatletter
\providecommand \@ifxundefined [1]{%
 \@ifx{#1\undefined}
}%
\providecommand \@ifnum [1]{%
 \ifnum #1\expandafter \@firstoftwo
 \else \expandafter \@secondoftwo
 \fi
}%
\providecommand \@ifx [1]{%
 \ifx #1\expandafter \@firstoftwo
 \else \expandafter \@secondoftwo
 \fi
}%
\providecommand \natexlab [1]{#1}%
\providecommand \enquote  [1]{``#1''}%
\providecommand \bibnamefont  [1]{#1}%
\providecommand \bibfnamefont [1]{#1}%
\providecommand \citenamefont [1]{#1}%
\providecommand \href@noop [0]{\@secondoftwo}%
\providecommand \href [0]{\begingroup \@sanitize@url \@href}%
\providecommand \@href[1]{\@@startlink{#1}\@@href}%
\providecommand \@@href[1]{\endgroup#1\@@endlink}%
\providecommand \@sanitize@url [0]{\catcode `\\12\catcode `\$12\catcode
  `\&12\catcode `\#12\catcode `\^12\catcode `\_12\catcode `\%12\relax}%
\providecommand \@@startlink[1]{}%
\providecommand \@@endlink[0]{}%
\providecommand \url  [0]{\begingroup\@sanitize@url \@url }%
\providecommand \@url [1]{\endgroup\@href {#1}{\urlprefix }}%
\providecommand \urlprefix  [0]{URL }%
\providecommand \Eprint [0]{\href }%
\providecommand \doibase [0]{http://dx.doi.org/}%
\providecommand \selectlanguage [0]{\@gobble}%
\providecommand \bibinfo  [0]{\@secondoftwo}%
\providecommand \bibfield  [0]{\@secondoftwo}%
\providecommand \translation [1]{[#1]}%
\providecommand \BibitemOpen [0]{}%
\providecommand \bibitemStop [0]{}%
\providecommand \bibitemNoStop [0]{.\EOS\space}%
\providecommand \EOS [0]{\spacefactor3000\relax}%
\providecommand \BibitemShut  [1]{\csname bibitem#1\endcsname}%
\let\auto@bib@innerbib\@empty
%</preamble>
\bibitem [{\citenamefont {Abrikosov}()}]{abrikosov_book_1988}%
  \BibitemOpen
  \bibfield  {author} {\bibinfo {author} {\bibfnamefont {A.}~\bibnamefont
  {Abrikosov}},\ }\href@noop {} {\emph {\bibinfo {title} {Fundamentals of the
  Theory of Metals}}}\BibitemShut {NoStop}%
\bibitem [{\citenamefont {Lifshitz}(1930)}]{lifshitz}%
  \BibitemOpen
  \bibfield  {author} {\bibinfo {author} {\bibfnamefont {I.~M.}\ \bibnamefont
  {Lifshitz}},\ }\href@noop {} {\bibfield  {journal} {\bibinfo  {journal} {Sov.
  Phys. JETP}\ }\textbf {\bibinfo {volume} {11}},\ \bibinfo {pages} {1130}
  (\bibinfo {year} {1930})}\BibitemShut {NoStop}%
\bibitem [{\citenamefont {Blanter}\ \emph {et~al.}(1994)\citenamefont
  {Blanter}, \citenamefont {Kaganov}, \citenamefont {Pantsulaya},\ and\
  \citenamefont {Varlamov}}]{blanter_physrep_1994}%
  \BibitemOpen
  \bibfield  {author} {\bibinfo {author} {\bibfnamefont {Y.~M.}\ \bibnamefont
  {Blanter}}, \bibinfo {author} {\bibfnamefont {M.}~\bibnamefont {Kaganov}},
  \bibinfo {author} {\bibfnamefont {A.}~\bibnamefont {Pantsulaya}}, \ and\
  \bibinfo {author} {\bibfnamefont {A.}~\bibnamefont {Varlamov}},\ }\href@noop
  {} {\bibfield  {journal} {\bibinfo  {journal} {Physics Reports}\ }\textbf
  {\bibinfo {volume} {245}},\ \bibinfo {pages} {159} (\bibinfo {year}
  {1994})}\BibitemShut {NoStop}%
\bibitem [{\citenamefont {Bruno}\ \emph {et~al.}(1994)\citenamefont {Bruno},
  \citenamefont {Ginatempo}, \citenamefont {Guiliano}, \citenamefont {Ruban},\
  and\ \citenamefont {Vekilov}}]{bruno_physrep_1994}%
  \BibitemOpen
  \bibfield  {author} {\bibinfo {author} {\bibfnamefont {E.}~\bibnamefont
  {Bruno}}, \bibinfo {author} {\bibfnamefont {B.}~\bibnamefont {Ginatempo}},
  \bibinfo {author} {\bibfnamefont {E.}~\bibnamefont {Guiliano}}, \bibinfo
  {author} {\bibfnamefont {A.}~\bibnamefont {Ruban}}, \ and\ \bibinfo {author}
  {\bibfnamefont {Y.~K.}\ \bibnamefont {Vekilov}},\ }\href@noop {} {\bibfield
  {journal} {\bibinfo  {journal} {Physics Reports}\ }\textbf {\bibinfo {volume}
  {249}},\ \bibinfo {pages} {353} (\bibinfo {year} {1994})}\BibitemShut
  {NoStop}%
\bibitem [{\citenamefont {Yoshizumi}\ \emph {et~al.}(2007)\citenamefont
  {Yoshizumi}, \citenamefont {Muraoka}, \citenamefont {Okamoto}, \citenamefont
  {Kiuchi}, \citenamefont {Yamaura}, \citenamefont {Mochizuki}, \citenamefont
  {Ogata},\ and\ \citenamefont {Hiroi}}]{yoshizumi_jpsj_2007}%
  \BibitemOpen
  \bibfield  {author} {\bibinfo {author} {\bibfnamefont {D.}~\bibnamefont
  {Yoshizumi}}, \bibinfo {author} {\bibfnamefont {Y.}~\bibnamefont {Muraoka}},
  \bibinfo {author} {\bibfnamefont {Y.}~\bibnamefont {Okamoto}}, \bibinfo
  {author} {\bibfnamefont {Y.}~\bibnamefont {Kiuchi}}, \bibinfo {author}
  {\bibfnamefont {J.-I.}\ \bibnamefont {Yamaura}}, \bibinfo {author}
  {\bibfnamefont {M.}~\bibnamefont {Mochizuki}}, \bibinfo {author}
  {\bibfnamefont {M.}~\bibnamefont {Ogata}}, \ and\ \bibinfo {author}
  {\bibfnamefont {Z.}~\bibnamefont {Hiroi}},\ }\href@noop {} {\bibfield
  {journal} {\bibinfo  {journal} {Journal of the Physical Society of Japan}\
  }\textbf {\bibinfo {volume} {76}} (\bibinfo {year} {2007})}\BibitemShut
  {NoStop}%
\bibitem [{\citenamefont {Okamoto}\ \emph {et~al.}(2010)\citenamefont
  {Okamoto}, \citenamefont {Nishio},\ and\ \citenamefont
  {Hiroi}}]{okamoto_prb_2010}%
  \BibitemOpen
  \bibfield  {author} {\bibinfo {author} {\bibfnamefont {Y.}~\bibnamefont
  {Okamoto}}, \bibinfo {author} {\bibfnamefont {A.}~\bibnamefont {Nishio}}, \
  and\ \bibinfo {author} {\bibfnamefont {Z.}~\bibnamefont {Hiroi}},\
  }\href@noop {} {\bibfield  {journal} {\bibinfo  {journal} {Physical Review
  B}\ }\textbf {\bibinfo {volume} {81}},\ \bibinfo {pages} {121102} (\bibinfo
  {year} {2010})}\BibitemShut {NoStop}%
\bibitem [{\citenamefont {Sebastian}\ \emph {et~al.}(2010)\citenamefont
  {Sebastian}, \citenamefont {Harrison}, \citenamefont {Altarawneh},
  \citenamefont {Mielke}, \citenamefont {Liang}, \citenamefont {Bonn},\ and\
  \citenamefont {Lonzarich}}]{sebastian_pnas_2010}%
  \BibitemOpen
  \bibfield  {author} {\bibinfo {author} {\bibfnamefont {S.~E.}\ \bibnamefont
  {Sebastian}}, \bibinfo {author} {\bibfnamefont {N.}~\bibnamefont {Harrison}},
  \bibinfo {author} {\bibfnamefont {M.}~\bibnamefont {Altarawneh}}, \bibinfo
  {author} {\bibfnamefont {C.}~\bibnamefont {Mielke}}, \bibinfo {author}
  {\bibfnamefont {R.}~\bibnamefont {Liang}}, \bibinfo {author} {\bibfnamefont
  {D.}~\bibnamefont {Bonn}}, \ and\ \bibinfo {author} {\bibfnamefont
  {G.}~\bibnamefont {Lonzarich}},\ }\href@noop {} {\bibfield  {journal}
  {\bibinfo  {journal} {Proceedings of the National Academy of Sciences}\
  }\textbf {\bibinfo {volume} {107}},\ \bibinfo {pages} {6175} (\bibinfo {year}
  {2010})}\BibitemShut {NoStop}%
\bibitem [{\citenamefont {Norman}\ \emph {et~al.}(2010)\citenamefont {Norman},
  \citenamefont {Lin},\ and\ \citenamefont {Millis}}]{norman_prb_2010}%
  \BibitemOpen
  \bibfield  {author} {\bibinfo {author} {\bibfnamefont {M.}~\bibnamefont
  {Norman}}, \bibinfo {author} {\bibfnamefont {J.}~\bibnamefont {Lin}}, \ and\
  \bibinfo {author} {\bibfnamefont {A.}~\bibnamefont {Millis}},\ }\href@noop {}
  {\bibfield  {journal} {\bibinfo  {journal} {Physical Review B}\ }\textbf
  {\bibinfo {volume} {81}},\ \bibinfo {pages} {180513} (\bibinfo {year}
  {2010})}\BibitemShut {NoStop}%
\bibitem [{\citenamefont {LeBoeuf}\ \emph {et~al.}(2011)\citenamefont
  {LeBoeuf}, \citenamefont {Doiron-Leyraud}, \citenamefont {Vignolle},
  \citenamefont {Sutherland}, \citenamefont {Ramshaw}, \citenamefont
  {Levallois}, \citenamefont {Daou}, \citenamefont {Lalibert{\'e}},
  \citenamefont {Cyr-Choiniere}, \citenamefont {Chang} \emph
  {et~al.}}]{leboeuf_prb_2011}%
  \BibitemOpen
  \bibfield  {author} {\bibinfo {author} {\bibfnamefont {D.}~\bibnamefont
  {LeBoeuf}}, \bibinfo {author} {\bibfnamefont {N.}~\bibnamefont
  {Doiron-Leyraud}}, \bibinfo {author} {\bibfnamefont {B.}~\bibnamefont
  {Vignolle}}, \bibinfo {author} {\bibfnamefont {M.}~\bibnamefont
  {Sutherland}}, \bibinfo {author} {\bibfnamefont {B.}~\bibnamefont {Ramshaw}},
  \bibinfo {author} {\bibfnamefont {J.}~\bibnamefont {Levallois}}, \bibinfo
  {author} {\bibfnamefont {R.}~\bibnamefont {Daou}}, \bibinfo {author}
  {\bibfnamefont {F.}~\bibnamefont {Lalibert{\'e}}}, \bibinfo {author}
  {\bibfnamefont {O.}~\bibnamefont {Cyr-Choiniere}}, \bibinfo {author}
  {\bibfnamefont {J.}~\bibnamefont {Chang}},  \emph {et~al.},\ }\href@noop {}
  {\bibfield  {journal} {\bibinfo  {journal} {Physical Review B}\ }\textbf
  {\bibinfo {volume} {83}},\ \bibinfo {pages} {054506} (\bibinfo {year}
  {2011})}\BibitemShut {NoStop}%
\bibitem [{\citenamefont {Chu}\ \emph {et~al.}(1970)\citenamefont {Chu},
  \citenamefont {Smith},\ and\ \citenamefont {Gardner}}]{chu_prb_1970}%
  \BibitemOpen
  \bibfield  {author} {\bibinfo {author} {\bibfnamefont {C.}~\bibnamefont
  {Chu}}, \bibinfo {author} {\bibfnamefont {T.}~\bibnamefont {Smith}}, \ and\
  \bibinfo {author} {\bibfnamefont {W.}~\bibnamefont {Gardner}},\ }\href@noop
  {} {\bibfield  {journal} {\bibinfo  {journal} {Physical Review B}\ }\textbf
  {\bibinfo {volume} {1}},\ \bibinfo {pages} {214} (\bibinfo {year}
  {1970})}\BibitemShut {NoStop}%
\bibitem [{\citenamefont {Godwal}\ \emph {et~al.}(1998)\citenamefont {Godwal},
  \citenamefont {Jayaraman}, \citenamefont {Meenakshi}, \citenamefont {Rao},
  \citenamefont {Sikka},\ and\ \citenamefont {Vijayakumar}}]{godwal_prb_1998}%
  \BibitemOpen
  \bibfield  {author} {\bibinfo {author} {\bibfnamefont {B.}~\bibnamefont
  {Godwal}}, \bibinfo {author} {\bibfnamefont {A.}~\bibnamefont {Jayaraman}},
  \bibinfo {author} {\bibfnamefont {S.}~\bibnamefont {Meenakshi}}, \bibinfo
  {author} {\bibfnamefont {R.}~\bibnamefont {Rao}}, \bibinfo {author}
  {\bibfnamefont {S.}~\bibnamefont {Sikka}}, \ and\ \bibinfo {author}
  {\bibfnamefont {V.}~\bibnamefont {Vijayakumar}},\ }\href@noop {} {\bibfield
  {journal} {\bibinfo  {journal} {Physical Review B}\ }\textbf {\bibinfo
  {volume} {57}},\ \bibinfo {pages} {773} (\bibinfo {year} {1998})}\BibitemShut
  {NoStop}%
\bibitem [{\citenamefont {Rourke}\ \emph {et~al.}(2008)\citenamefont {Rourke},
  \citenamefont {McCollam}, \citenamefont {Lapertot}, \citenamefont {Knebel},
  \citenamefont {Flouquet},\ and\ \citenamefont {Julian}}]{rourke_prl_2008}%
  \BibitemOpen
  \bibfield  {author} {\bibinfo {author} {\bibfnamefont {P.}~\bibnamefont
  {Rourke}}, \bibinfo {author} {\bibfnamefont {A.}~\bibnamefont {McCollam}},
  \bibinfo {author} {\bibfnamefont {G.}~\bibnamefont {Lapertot}}, \bibinfo
  {author} {\bibfnamefont {G.}~\bibnamefont {Knebel}}, \bibinfo {author}
  {\bibfnamefont {J.}~\bibnamefont {Flouquet}}, \ and\ \bibinfo {author}
  {\bibfnamefont {S.}~\bibnamefont {Julian}},\ }\href@noop {} {\bibfield
  {journal} {\bibinfo  {journal} {Physical review letters}\ }\textbf {\bibinfo
  {volume} {101}},\ \bibinfo {pages} {237205} (\bibinfo {year}
  {2008})}\BibitemShut {NoStop}%
\bibitem [{\citenamefont {Wosnitza}\ \emph {et~al.}(2008)\citenamefont
  {Wosnitza}, \citenamefont {Goll}, \citenamefont {Bartkowiak}, \citenamefont
  {Bergk}, \citenamefont {Bianchi}, \citenamefont {L{\"o}hneysen},
  \citenamefont {Yoshino},\ and\ \citenamefont
  {Takabatake}}]{wosnitza_physb_2008}%
  \BibitemOpen
  \bibfield  {author} {\bibinfo {author} {\bibfnamefont {J.}~\bibnamefont
  {Wosnitza}}, \bibinfo {author} {\bibfnamefont {G.}~\bibnamefont {Goll}},
  \bibinfo {author} {\bibfnamefont {M.}~\bibnamefont {Bartkowiak}}, \bibinfo
  {author} {\bibfnamefont {B.}~\bibnamefont {Bergk}}, \bibinfo {author}
  {\bibfnamefont {A.}~\bibnamefont {Bianchi}}, \bibinfo {author} {\bibfnamefont
  {H.}~\bibnamefont {L{\"o}hneysen}}, \bibinfo {author} {\bibfnamefont
  {T.}~\bibnamefont {Yoshino}}, \ and\ \bibinfo {author} {\bibfnamefont
  {T.}~\bibnamefont {Takabatake}},\ }\href@noop {} {\bibfield  {journal}
  {\bibinfo  {journal} {Physica B: Condensed Matter}\ }\textbf {\bibinfo
  {volume} {403}},\ \bibinfo {pages} {1219} (\bibinfo {year}
  {2008})}\BibitemShut {NoStop}%
\bibitem [{\citenamefont {{McCann}}\ and\ \citenamefont
  {Fal’ko}(2006)}]{mccann_landau-level_2006}%
  \BibitemOpen
  \bibfield  {author} {\bibinfo {author} {\bibfnamefont {E.}~\bibnamefont
  {{McCann}}}\ and\ \bibinfo {author} {\bibfnamefont {V.}~\bibnamefont
  {Fal’ko}},\ }\href {\doibase 10.1103/PhysRevLett.96.086805} {\bibfield
  {journal} {\bibinfo  {journal} {Physical Review Letters}\ }\textbf {\bibinfo
  {volume} {96}},\ \bibinfo {pages} {086805} (\bibinfo {year}
  {2006})}\BibitemShut {NoStop}%
\bibitem [{\citenamefont {McClure}(1957)}]{mcclure1957}%
  \BibitemOpen
  \bibfield  {author} {\bibinfo {author} {\bibfnamefont {J.~W.}\ \bibnamefont
  {McClure}},\ }\href {\doibase 10.1103/PhysRev.108.612} {\bibfield  {journal}
  {\bibinfo  {journal} {Phys. Rev.}\ }\textbf {\bibinfo {volume} {108}},\
  \bibinfo {pages} {612} (\bibinfo {year} {1957})}\BibitemShut {NoStop}%
\bibitem [{\citenamefont {Inoue}(1962)}]{inoue_jpsj_1962}%
  \BibitemOpen
  \bibfield  {author} {\bibinfo {author} {\bibfnamefont {M.}~\bibnamefont
  {Inoue}},\ }\href@noop {} {\bibfield  {journal} {\bibinfo  {journal} {Journal
  of the Physical Society of Japan}\ }\textbf {\bibinfo {volume} {17}},\
  \bibinfo {pages} {808} (\bibinfo {year} {1962})}\BibitemShut {NoStop}%
\bibitem [{\citenamefont {Williamson}\ \emph {et~al.}(1966)\citenamefont
  {Williamson}, \citenamefont {Surma}, \citenamefont {Pradduade}, \citenamefont
  {Patten},\ and\ \citenamefont {Furdyna}}]{williamson_ssc_1966}%
  \BibitemOpen
  \bibfield  {author} {\bibinfo {author} {\bibfnamefont {S.}~\bibnamefont
  {Williamson}}, \bibinfo {author} {\bibfnamefont {M.}~\bibnamefont {Surma}},
  \bibinfo {author} {\bibfnamefont {H.}~\bibnamefont {Pradduade}}, \bibinfo
  {author} {\bibfnamefont {R.}~\bibnamefont {Patten}}, \ and\ \bibinfo {author}
  {\bibfnamefont {J.}~\bibnamefont {Furdyna}},\ }\href@noop {} {\bibfield
  {journal} {\bibinfo  {journal} {Solid State Communications}\ }\textbf
  {\bibinfo {volume} {4}},\ \bibinfo {pages} {37} (\bibinfo {year}
  {1966})}\BibitemShut {NoStop}%
\bibitem [{\citenamefont {Orlita}\ \emph {et~al.}(2012)\citenamefont {Orlita},
  \citenamefont {Neugebauer}, \citenamefont {Faugeras}, \citenamefont {Barra},
  \citenamefont {Potemski}, \citenamefont {Pellegrino},\ and\ \citenamefont
  {Basko}}]{orlita_prl_2012}%
  \BibitemOpen
  \bibfield  {author} {\bibinfo {author} {\bibfnamefont {M.}~\bibnamefont
  {Orlita}}, \bibinfo {author} {\bibfnamefont {P.}~\bibnamefont {Neugebauer}},
  \bibinfo {author} {\bibfnamefont {C.}~\bibnamefont {Faugeras}}, \bibinfo
  {author} {\bibfnamefont {A.-L.}\ \bibnamefont {Barra}}, \bibinfo {author}
  {\bibfnamefont {M.}~\bibnamefont {Potemski}}, \bibinfo {author}
  {\bibfnamefont {F.}~\bibnamefont {Pellegrino}}, \ and\ \bibinfo {author}
  {\bibfnamefont {D.}~\bibnamefont {Basko}},\ }\href@noop {} {\bibfield
  {journal} {\bibinfo  {journal} {Physical review letters}\ }\textbf {\bibinfo
  {volume} {108}},\ \bibinfo {pages} {017602} (\bibinfo {year}
  {2012})}\BibitemShut {NoStop}%
\bibitem [{\citenamefont {Mucha-Kruczy{\'n}ski}\ \emph
  {et~al.}(2011{\natexlab{a}})\citenamefont {Mucha-Kruczy{\'n}ski},
  \citenamefont {Aleiner},\ and\ \citenamefont
  {Fal'ko}}]{mucha-kruczynski_prb_2011}%
  \BibitemOpen
  \bibfield  {author} {\bibinfo {author} {\bibfnamefont {M.}~\bibnamefont
  {Mucha-Kruczy{\'n}ski}}, \bibinfo {author} {\bibfnamefont {I.~L.}\
  \bibnamefont {Aleiner}}, \ and\ \bibinfo {author} {\bibfnamefont {V.~I.}\
  \bibnamefont {Fal'ko}},\ }\href@noop {} {\bibfield  {journal} {\bibinfo
  {journal} {Physical Review B}\ }\textbf {\bibinfo {volume} {84}},\ \bibinfo
  {pages} {041404} (\bibinfo {year} {2011}{\natexlab{a}})}\BibitemShut
  {NoStop}%
\bibitem [{\citenamefont {Mucha-Kruczy{\'n}ski}\ \emph
  {et~al.}(2011{\natexlab{b}})\citenamefont {Mucha-Kruczy{\'n}ski},
  \citenamefont {Aleiner},\ and\ \citenamefont
  {Fal’ko}}]{mucha-kruczynski_ssc_2011}%
  \BibitemOpen
  \bibfield  {author} {\bibinfo {author} {\bibfnamefont {M.}~\bibnamefont
  {Mucha-Kruczy{\'n}ski}}, \bibinfo {author} {\bibfnamefont {I.~L.}\
  \bibnamefont {Aleiner}}, \ and\ \bibinfo {author} {\bibfnamefont {V.~I.}\
  \bibnamefont {Fal’ko}},\ }\href@noop {} {\bibfield  {journal} {\bibinfo
  {journal} {Solid State Communications}\ }\textbf {\bibinfo {volume} {151}},\
  \bibinfo {pages} {1088} (\bibinfo {year} {2011}{\natexlab{b}})}\BibitemShut
  {NoStop}%
\bibitem [{\citenamefont {Varlet}\ \emph
  {et~al.}(2014{\natexlab{a}})\citenamefont {Varlet}, \citenamefont {Bischoff},
  \citenamefont {Simonet}, \citenamefont {Watanabe}, \citenamefont {Taniguchi},
  \citenamefont {Ihn}, \citenamefont {Ensslin}, \citenamefont
  {Mucha-Kruczy\'{n}ski},\ and\ \citenamefont
  {Fal'ko}}]{varlet_anomalous_2014}%
  \BibitemOpen
  \bibfield  {author} {\bibinfo {author} {\bibfnamefont {A.}~\bibnamefont
  {Varlet}}, \bibinfo {author} {\bibfnamefont {D.}~\bibnamefont {Bischoff}},
  \bibinfo {author} {\bibfnamefont {P.}~\bibnamefont {Simonet}}, \bibinfo
  {author} {\bibfnamefont {K.}~\bibnamefont {Watanabe}}, \bibinfo {author}
  {\bibfnamefont {T.}~\bibnamefont {Taniguchi}}, \bibinfo {author}
  {\bibfnamefont {T.}~\bibnamefont {Ihn}}, \bibinfo {author} {\bibfnamefont
  {K.}~\bibnamefont {Ensslin}}, \bibinfo {author} {\bibfnamefont
  {M.}~\bibnamefont {Mucha-Kruczy\'{n}ski}}, \ and\ \bibinfo {author}
  {\bibfnamefont {V.}~\bibnamefont {Fal'ko}},\ }\href {\doibase
  10.1103/PhysRevLett.113.116602} {\bibfield  {journal} {\bibinfo  {journal}
  {Phys. Rev. Lett.}\ }\textbf {\bibinfo {volume} {113}},\ \bibinfo {pages}
  {116602} (\bibinfo {year} {2014}{\natexlab{a}})}\BibitemShut {NoStop}%
\bibitem [{\citenamefont {Novoselov}\ \emph {et~al.}(2004)\citenamefont
  {Novoselov}, \citenamefont {Geim}, \citenamefont {Morozov}, \citenamefont
  {Jiang}, \citenamefont {Zhang}, \citenamefont {Dubonos}, \citenamefont
  {Grigorieva},\ and\ \citenamefont {Firsov}}]{novoselov2004}%
  \BibitemOpen
  \bibfield  {author} {\bibinfo {author} {\bibfnamefont {K.~S.}\ \bibnamefont
  {Novoselov}}, \bibinfo {author} {\bibfnamefont {A.~K.}\ \bibnamefont {Geim}},
  \bibinfo {author} {\bibfnamefont {S.}~\bibnamefont {Morozov}}, \bibinfo
  {author} {\bibfnamefont {D.}~\bibnamefont {Jiang}}, \bibinfo {author}
  {\bibfnamefont {Y.}~\bibnamefont {Zhang}}, \bibinfo {author} {\bibfnamefont
  {S.}~\bibnamefont {Dubonos}}, \bibinfo {author} {\bibfnamefont
  {I.}~\bibnamefont {Grigorieva}}, \ and\ \bibinfo {author} {\bibfnamefont
  {A.}~\bibnamefont {Firsov}},\ }\href {\doibase 10.1126/science.1102896}
  {\bibfield  {journal} {\bibinfo  {journal} {Science}\ }\textbf {\bibinfo
  {volume} {306}},\ \bibinfo {pages} {666} (\bibinfo {year}
  {2004})}\BibitemShut {NoStop}%
\bibitem [{\citenamefont {Geim}(2009)}]{geim_science_2009}%
  \BibitemOpen
  \bibfield  {author} {\bibinfo {author} {\bibfnamefont {A.~K.}\ \bibnamefont
  {Geim}},\ }\href@noop {} {\bibfield  {journal} {\bibinfo  {journal}
  {Science}\ }\textbf {\bibinfo {volume} {324}},\ \bibinfo {pages} {1530}
  (\bibinfo {year} {2009})}\BibitemShut {NoStop}%
\bibitem [{\citenamefont {Kuzmenko}\ \emph {et~al.}(2009)\citenamefont
  {Kuzmenko}, \citenamefont {Crassee}, \citenamefont {Van Der~Marel},
  \citenamefont {Blake},\ and\ \citenamefont {Novoselov}}]{kuzmenko_prb_2009}%
  \BibitemOpen
  \bibfield  {author} {\bibinfo {author} {\bibfnamefont {A.}~\bibnamefont
  {Kuzmenko}}, \bibinfo {author} {\bibfnamefont {I.}~\bibnamefont {Crassee}},
  \bibinfo {author} {\bibfnamefont {D.}~\bibnamefont {Van Der~Marel}}, \bibinfo
  {author} {\bibfnamefont {P.}~\bibnamefont {Blake}}, \ and\ \bibinfo {author}
  {\bibfnamefont {K.}~\bibnamefont {Novoselov}},\ }\href@noop {} {\bibfield
  {journal} {\bibinfo  {journal} {Physical Review B}\ }\textbf {\bibinfo
  {volume} {80}},\ \bibinfo {pages} {165406} (\bibinfo {year}
  {2009})}\BibitemShut {NoStop}%
\bibitem [{\citenamefont {Wallace}(1947)}]{wallace_physrev_1947}%
  \BibitemOpen
  \bibfield  {author} {\bibinfo {author} {\bibfnamefont {P.~R.}\ \bibnamefont
  {Wallace}},\ }\href@noop {} {\bibfield  {journal} {\bibinfo  {journal}
  {Physical Review}\ }\textbf {\bibinfo {volume} {71}},\ \bibinfo {pages} {622}
  (\bibinfo {year} {1947})}\BibitemShut {NoStop}%
\bibitem [{\citenamefont {Novoselov}\ \emph {et~al.}(2006)\citenamefont
  {Novoselov}, \citenamefont {{McCann}}, \citenamefont {Morozov}, \citenamefont
  {Fal’ko}, \citenamefont {Katsnelson}, \citenamefont {Zeitler},
  \citenamefont {Jiang}, \citenamefont {Schedin},\ and\ \citenamefont
  {Geim}}]{novoselov2006}%
  \BibitemOpen
  \bibfield  {author} {\bibinfo {author} {\bibfnamefont {K.~S.}\ \bibnamefont
  {Novoselov}}, \bibinfo {author} {\bibfnamefont {E.}~\bibnamefont {{McCann}}},
  \bibinfo {author} {\bibfnamefont {S.~V.}\ \bibnamefont {Morozov}}, \bibinfo
  {author} {\bibfnamefont {V.~I.}\ \bibnamefont {Fal’ko}}, \bibinfo {author}
  {\bibfnamefont {M.~I.}\ \bibnamefont {Katsnelson}}, \bibinfo {author}
  {\bibfnamefont {U.}~\bibnamefont {Zeitler}}, \bibinfo {author} {\bibfnamefont
  {D.}~\bibnamefont {Jiang}}, \bibinfo {author} {\bibfnamefont
  {F.}~\bibnamefont {Schedin}}, \ and\ \bibinfo {author} {\bibfnamefont
  {A.~K.}\ \bibnamefont {Geim}},\ }\href {\doibase 10.1038/nphys245} {\bibfield
   {journal} {\bibinfo  {journal} {Nature Physics}\ }\textbf {\bibinfo {volume}
  {2}},\ \bibinfo {pages} {177} (\bibinfo {year} {2006})}\BibitemShut {NoStop}%
\bibitem [{\citenamefont {Cohen}\ and\ \citenamefont
  {Falicov}(1961)}]{cohen_prl_1961}%
  \BibitemOpen
  \bibfield  {author} {\bibinfo {author} {\bibfnamefont {M.~H.}\ \bibnamefont
  {Cohen}}\ and\ \bibinfo {author} {\bibfnamefont {L.}~\bibnamefont
  {Falicov}},\ }\href@noop {} {\bibfield  {journal} {\bibinfo  {journal}
  {Physical Review Letters}\ }\textbf {\bibinfo {volume} {7}},\ \bibinfo
  {pages} {231} (\bibinfo {year} {1961})}\BibitemShut {NoStop}%
\bibitem [{\citenamefont {Blount}(1962)}]{blount_physrev_1962}%
  \BibitemOpen
  \bibfield  {author} {\bibinfo {author} {\bibfnamefont {E.}~\bibnamefont
  {Blount}},\ }\href@noop {} {\bibfield  {journal} {\bibinfo  {journal}
  {Physical Review}\ }\textbf {\bibinfo {volume} {126}},\ \bibinfo {pages}
  {1636} (\bibinfo {year} {1962})}\BibitemShut {NoStop}%
\bibitem [{\citenamefont {Novoselov}\ \emph {et~al.}(2005)\citenamefont
  {Novoselov}, \citenamefont {Geim}, \citenamefont {Morozov}, \citenamefont
  {Jiang}, \citenamefont {Grigorieva}, \citenamefont {Dubonos},\ and\
  \citenamefont {Firsov}}]{novoselov2005}%
  \BibitemOpen
  \bibfield  {author} {\bibinfo {author} {\bibfnamefont {K.}~\bibnamefont
  {Novoselov}}, \bibinfo {author} {\bibfnamefont {A.~K.}\ \bibnamefont {Geim}},
  \bibinfo {author} {\bibfnamefont {S.}~\bibnamefont {Morozov}}, \bibinfo
  {author} {\bibfnamefont {D.}~\bibnamefont {Jiang}}, \bibinfo {author}
  {\bibfnamefont {M.~K.~I.}\ \bibnamefont {Grigorieva}}, \bibinfo {author}
  {\bibfnamefont {S.}~\bibnamefont {Dubonos}}, \ and\ \bibinfo {author}
  {\bibfnamefont {A.}~\bibnamefont {Firsov}},\ }\href {\doibase
  10.1038/nature04233} {\bibfield  {journal} {\bibinfo  {journal} {Nature}\
  }\textbf {\bibinfo {volume} {438}},\ \bibinfo {pages} {197} (\bibinfo {year}
  {2005})}\BibitemShut {NoStop}%
\bibitem [{\citenamefont {Zhang}\ \emph {et~al.}(2005)\citenamefont {Zhang},
  \citenamefont {Tan}, \citenamefont {Stormer},\ and\ \citenamefont
  {Kim}}]{zhang2005}%
  \BibitemOpen
  \bibfield  {author} {\bibinfo {author} {\bibfnamefont {Y.}~\bibnamefont
  {Zhang}}, \bibinfo {author} {\bibfnamefont {Y.-W.}\ \bibnamefont {Tan}},
  \bibinfo {author} {\bibfnamefont {H.~L.}\ \bibnamefont {Stormer}}, \ and\
  \bibinfo {author} {\bibfnamefont {P.}~\bibnamefont {Kim}},\ }\href {\doibase
  10.1038/nature04235} {\bibfield  {journal} {\bibinfo  {journal} {Nature}\
  }\textbf {\bibinfo {volume} {438}},\ \bibinfo {pages} {201} (\bibinfo {year}
  {2005})}\BibitemShut {NoStop}%
\bibitem [{\citenamefont {Ando}(2006)}]{ando2006}%
  \BibitemOpen
  \bibfield  {author} {\bibinfo {author} {\bibfnamefont {T.}~\bibnamefont
  {Ando}},\ }\href {\doibase 10.1143/JPSJ.75.074716} {\bibfield  {journal}
  {\bibinfo  {journal} {Journal of the Physical Society of Japan}\ }\textbf
  {\bibinfo {volume} {75}},\ \bibinfo {pages} {074716} (\bibinfo {year}
  {2006})}\BibitemShut {NoStop}%
\bibitem [{\citenamefont {Ishigami}\ \emph {et~al.}(2007)\citenamefont
  {Ishigami}, \citenamefont {Chen}, \citenamefont {Cullen}, \citenamefont
  {Fuhrer},\ and\ \citenamefont {Williams}}]{ishigami2007}%
  \BibitemOpen
  \bibfield  {author} {\bibinfo {author} {\bibfnamefont {M.}~\bibnamefont
  {Ishigami}}, \bibinfo {author} {\bibfnamefont {J.~H.}\ \bibnamefont {Chen}},
  \bibinfo {author} {\bibfnamefont {W.~G.}\ \bibnamefont {Cullen}}, \bibinfo
  {author} {\bibfnamefont {M.~S.}\ \bibnamefont {Fuhrer}}, \ and\ \bibinfo
  {author} {\bibfnamefont {E.~D.}\ \bibnamefont {Williams}},\ }\href {\doibase
  10.1021/nl070613a} {\bibfield  {journal} {\bibinfo  {journal} {Nano Letters}\
  }\textbf {\bibinfo {volume} {7}},\ \bibinfo {pages} {1643} (\bibinfo {year}
  {2007})}\BibitemShut {NoStop}%
\bibitem [{\citenamefont {Fratini}\ and\ \citenamefont
  {Guinea}(2008)}]{fratini2008}%
  \BibitemOpen
  \bibfield  {author} {\bibinfo {author} {\bibfnamefont {S.}~\bibnamefont
  {Fratini}}\ and\ \bibinfo {author} {\bibfnamefont {F.}~\bibnamefont
  {Guinea}},\ }\href {\doibase 10.1103/PhysRevB.77.195415} {\bibfield
  {journal} {\bibinfo  {journal} {Phys. Rev. B}\ }\textbf {\bibinfo {volume}
  {77}},\ \bibinfo {pages} {195415} (\bibinfo {year} {2008})}\BibitemShut
  {NoStop}%
\bibitem [{\citenamefont {Chen}\ \emph {et~al.}(2008)\citenamefont {Chen},
  \citenamefont {Jang}, \citenamefont {Xiao}, \citenamefont {Ishigami},\ and\
  \citenamefont {Fuhrer}}]{chen2008}%
  \BibitemOpen
  \bibfield  {author} {\bibinfo {author} {\bibfnamefont {J.-H.}\ \bibnamefont
  {Chen}}, \bibinfo {author} {\bibfnamefont {C.}~\bibnamefont {Jang}}, \bibinfo
  {author} {\bibfnamefont {S.}~\bibnamefont {Xiao}}, \bibinfo {author}
  {\bibfnamefont {M.}~\bibnamefont {Ishigami}}, \ and\ \bibinfo {author}
  {\bibfnamefont {M.~S.}\ \bibnamefont {Fuhrer}},\ }\href {\doibase
  10.1038/nnano.2008.58} {\bibfield  {journal} {\bibinfo  {journal} {Nat Nano}\
  }\textbf {\bibinfo {volume} {3}},\ \bibinfo {pages} {206} (\bibinfo {year}
  {2008})}\BibitemShut {NoStop}%
\bibitem [{\citenamefont {Bolotin}\ \emph {et~al.}(2008)\citenamefont
  {Bolotin}, \citenamefont {Sikes}, \citenamefont {Hone}, \citenamefont
  {Stormer},\ and\ \citenamefont {Kim}}]{bolotin2008}%
  \BibitemOpen
  \bibfield  {author} {\bibinfo {author} {\bibfnamefont {K.~I.}\ \bibnamefont
  {Bolotin}}, \bibinfo {author} {\bibfnamefont {K.~J.}\ \bibnamefont {Sikes}},
  \bibinfo {author} {\bibfnamefont {J.}~\bibnamefont {Hone}}, \bibinfo {author}
  {\bibfnamefont {H.~L.}\ \bibnamefont {Stormer}}, \ and\ \bibinfo {author}
  {\bibfnamefont {P.}~\bibnamefont {Kim}},\ }\href {\doibase
  10.1103/PhysRevLett.101.096802} {\bibfield  {journal} {\bibinfo  {journal}
  {Phys. Rev. Lett.}\ }\textbf {\bibinfo {volume} {101}},\ \bibinfo {pages}
  {096802} (\bibinfo {year} {2008})}\BibitemShut {NoStop}%
\bibitem [{\citenamefont {Du}\ \emph {et~al.}(2008)\citenamefont {Du},
  \citenamefont {Skachko}, \citenamefont {Barker},\ and\ \citenamefont
  {Andrei}}]{du2008}%
  \BibitemOpen
  \bibfield  {author} {\bibinfo {author} {\bibfnamefont {X.}~\bibnamefont
  {Du}}, \bibinfo {author} {\bibfnamefont {I.}~\bibnamefont {Skachko}},
  \bibinfo {author} {\bibfnamefont {A.}~\bibnamefont {Barker}}, \ and\ \bibinfo
  {author} {\bibfnamefont {E.~Y.}\ \bibnamefont {Andrei}},\ }\href {\doibase
  10.1038/nnano.2008.199} {\bibfield  {journal} {\bibinfo  {journal} {Nat
  Nano}\ }\textbf {\bibinfo {volume} {3}},\ \bibinfo {pages} {491} (\bibinfo
  {year} {2008})}\BibitemShut {NoStop}%
\bibitem [{\citenamefont {Zhang}\ \emph {et~al.}(2014)\citenamefont {Zhang},
  \citenamefont {Huang}, \citenamefont {Velasco~Jr}, \citenamefont {Myhro},
  \citenamefont {Maldonado}, \citenamefont {Tran}, \citenamefont {Zhao},
  \citenamefont {Wang}, \citenamefont {Lee}, \citenamefont {Liu} \emph
  {et~al.}}]{zhang_carbon_2014}%
  \BibitemOpen
  \bibfield  {author} {\bibinfo {author} {\bibfnamefont {H.}~\bibnamefont
  {Zhang}}, \bibinfo {author} {\bibfnamefont {J.-W.}\ \bibnamefont {Huang}},
  \bibinfo {author} {\bibfnamefont {J.}~\bibnamefont {Velasco~Jr}}, \bibinfo
  {author} {\bibfnamefont {K.}~\bibnamefont {Myhro}}, \bibinfo {author}
  {\bibfnamefont {M.}~\bibnamefont {Maldonado}}, \bibinfo {author}
  {\bibfnamefont {D.~D.}\ \bibnamefont {Tran}}, \bibinfo {author}
  {\bibfnamefont {Z.}~\bibnamefont {Zhao}}, \bibinfo {author} {\bibfnamefont
  {F.}~\bibnamefont {Wang}}, \bibinfo {author} {\bibfnamefont {Y.}~\bibnamefont
  {Lee}}, \bibinfo {author} {\bibfnamefont {G.}~\bibnamefont {Liu}},  \emph
  {et~al.},\ }\href@noop {} {\bibfield  {journal} {\bibinfo  {journal}
  {Carbon}\ }\textbf {\bibinfo {volume} {69}},\ \bibinfo {pages} {336}
  (\bibinfo {year} {2014})}\BibitemShut {NoStop}%
\bibitem [{\citenamefont {Dean}\ \emph {et~al.}(2010)\citenamefont {Dean},
  \citenamefont {Young}, \citenamefont {Meric}, \citenamefont {Lee},
  \citenamefont {Wang}, \citenamefont {Sorgenfrei}, \citenamefont {Watanabe},
  \citenamefont {Taniguchi}, \citenamefont {Kim}, \citenamefont {Shepard},\
  and\ \citenamefont {Hone}}]{dean2010}%
  \BibitemOpen
  \bibfield  {author} {\bibinfo {author} {\bibfnamefont {C.~R.}\ \bibnamefont
  {Dean}}, \bibinfo {author} {\bibfnamefont {A.~F.}\ \bibnamefont {Young}},
  \bibinfo {author} {\bibfnamefont {I.}~\bibnamefont {Meric}}, \bibinfo
  {author} {\bibfnamefont {C.}~\bibnamefont {Lee}}, \bibinfo {author}
  {\bibfnamefont {L.}~\bibnamefont {Wang}}, \bibinfo {author} {\bibfnamefont
  {S.}~\bibnamefont {Sorgenfrei}}, \bibinfo {author} {\bibfnamefont
  {K.}~\bibnamefont {Watanabe}}, \bibinfo {author} {\bibfnamefont
  {T.}~\bibnamefont {Taniguchi}}, \bibinfo {author} {\bibfnamefont
  {P.}~\bibnamefont {Kim}}, \bibinfo {author} {\bibfnamefont {K.~L.}\
  \bibnamefont {Shepard}}, \ and\ \bibinfo {author} {\bibfnamefont
  {J.}~\bibnamefont {Hone}},\ }\href {\doibase 10.1038/nnano.2010.172}
  {\bibfield  {journal} {\bibinfo  {journal} {Nature Nanotechnology}\ }\textbf
  {\bibinfo {volume} {5}},\ \bibinfo {pages} {722} (\bibinfo {year}
  {2010})}\BibitemShut {NoStop}%
\bibitem [{\citenamefont {Wang}\ \emph {et~al.}(2013)\citenamefont {Wang},
  \citenamefont {Meric}, \citenamefont {Huang}, \citenamefont {Gao},
  \citenamefont {Gao}, \citenamefont {Tran}, \citenamefont {Taniguchi},
  \citenamefont {Watanabe}, \citenamefont {Campos}, \citenamefont {Muller},
  \citenamefont {Guo}, \citenamefont {Kim}, \citenamefont {Hone}, \citenamefont
  {Shepard},\ and\ \citenamefont {Dean}}]{wang2013}%
  \BibitemOpen
  \bibfield  {author} {\bibinfo {author} {\bibfnamefont {L.}~\bibnamefont
  {Wang}}, \bibinfo {author} {\bibfnamefont {I.}~\bibnamefont {Meric}},
  \bibinfo {author} {\bibfnamefont {P.~Y.}\ \bibnamefont {Huang}}, \bibinfo
  {author} {\bibfnamefont {Q.}~\bibnamefont {Gao}}, \bibinfo {author}
  {\bibfnamefont {Y.}~\bibnamefont {Gao}}, \bibinfo {author} {\bibfnamefont
  {H.}~\bibnamefont {Tran}}, \bibinfo {author} {\bibfnamefont {T.}~\bibnamefont
  {Taniguchi}}, \bibinfo {author} {\bibfnamefont {K.}~\bibnamefont {Watanabe}},
  \bibinfo {author} {\bibfnamefont {L.~M.}\ \bibnamefont {Campos}}, \bibinfo
  {author} {\bibfnamefont {D.~A.}\ \bibnamefont {Muller}}, \bibinfo {author}
  {\bibfnamefont {J.}~\bibnamefont {Guo}}, \bibinfo {author} {\bibfnamefont
  {P.}~\bibnamefont {Kim}}, \bibinfo {author} {\bibfnamefont {J.}~\bibnamefont
  {Hone}}, \bibinfo {author} {\bibfnamefont {K.~L.}\ \bibnamefont {Shepard}}, \
  and\ \bibinfo {author} {\bibfnamefont {C.~R.}\ \bibnamefont {Dean}},\ }\href
  {\doibase 10.1126/science.1244358} {\bibfield  {journal} {\bibinfo  {journal}
  {Science}\ }\textbf {\bibinfo {volume} {342}},\ \bibinfo {pages} {614}
  (\bibinfo {year} {2013})}\BibitemShut {NoStop}%
\bibitem [{\citenamefont {Zomer}\ \emph {et~al.}(2014)\citenamefont {Zomer},
  \citenamefont {Guimarães}, \citenamefont {Brant}, \citenamefont {Tombros},\
  and\ \citenamefont {van Wees}}]{zomer2014}%
  \BibitemOpen
  \bibfield  {author} {\bibinfo {author} {\bibfnamefont {P.~J.}\ \bibnamefont
  {Zomer}}, \bibinfo {author} {\bibfnamefont {M.~H.~D.}\ \bibnamefont
  {Guimarães}}, \bibinfo {author} {\bibfnamefont {J.~C.}\ \bibnamefont
  {Brant}}, \bibinfo {author} {\bibfnamefont {N.}~\bibnamefont {Tombros}}, \
  and\ \bibinfo {author} {\bibfnamefont {B.~J.}\ \bibnamefont {van Wees}},\
  }\href {http://arxiv.org/abs/1403.0399} {\bibfield  {journal} {\bibinfo
  {journal} {{arXiv}:1403.0399 [cond-mat]}\ } (\bibinfo {year}
  {2014})}\BibitemShut {NoStop}%
\bibitem [{\citenamefont {Bischoff}\ \emph {et~al.}(2012)\citenamefont
  {Bischoff}, \citenamefont {Kr\"{a}henmann}, \citenamefont {Dr\"{o}scher},
  \citenamefont {Gruner}, \citenamefont {Barraud}, \citenamefont {Ihn},\ and\
  \citenamefont {Ensslin}}]{bischoff2012}%
  \BibitemOpen
  \bibfield  {author} {\bibinfo {author} {\bibfnamefont {D.}~\bibnamefont
  {Bischoff}}, \bibinfo {author} {\bibfnamefont {T.}~\bibnamefont
  {Kr\"{a}henmann}}, \bibinfo {author} {\bibfnamefont {S.}~\bibnamefont
  {Dr\"{o}scher}}, \bibinfo {author} {\bibfnamefont {M.~A.}\ \bibnamefont
  {Gruner}}, \bibinfo {author} {\bibfnamefont {C.}~\bibnamefont {Barraud}},
  \bibinfo {author} {\bibfnamefont {T.}~\bibnamefont {Ihn}}, \ and\ \bibinfo
  {author} {\bibfnamefont {K.}~\bibnamefont {Ensslin}},\ }\href {\doibase
  10.1063/1.4765345} {\bibfield  {journal} {\bibinfo  {journal} {Applied
  Physics Letters}\ }\textbf {\bibinfo {volume} {101}},\ \bibinfo {pages}
  {203103} (\bibinfo {year} {2012})}\BibitemShut {NoStop}%
\bibitem [{\citenamefont {Bischoff}\ \emph {et~al.}(2014)\citenamefont
  {Bischoff}, \citenamefont {Libisch}, \citenamefont {Burgd\"orfer},
  \citenamefont {Ihn},\ and\ \citenamefont {Ensslin}}]{bischoff2014}%
  \BibitemOpen
  \bibfield  {author} {\bibinfo {author} {\bibfnamefont {D.}~\bibnamefont
  {Bischoff}}, \bibinfo {author} {\bibfnamefont {F.}~\bibnamefont {Libisch}},
  \bibinfo {author} {\bibfnamefont {J.}~\bibnamefont {Burgd\"orfer}}, \bibinfo
  {author} {\bibfnamefont {T.}~\bibnamefont {Ihn}}, \ and\ \bibinfo {author}
  {\bibfnamefont {K.}~\bibnamefont {Ensslin}},\ }\href {\doibase
  10.1103/PhysRevB.90.115405} {\bibfield  {journal} {\bibinfo  {journal} {Phys.
  Rev. B}\ }\textbf {\bibinfo {volume} {90}},\ \bibinfo {pages} {115405}
  (\bibinfo {year} {2014})}\BibitemShut {NoStop}%
\bibitem [{\citenamefont {De~Gail}\ \emph {et~al.}(2011)\citenamefont
  {De~Gail}, \citenamefont {Goerbig}, \citenamefont {Guinea}, \citenamefont
  {Montambaux},\ and\ \citenamefont {Neto}}]{de_gail_prb_2011}%
  \BibitemOpen
  \bibfield  {author} {\bibinfo {author} {\bibfnamefont {R.}~\bibnamefont
  {De~Gail}}, \bibinfo {author} {\bibfnamefont {M.}~\bibnamefont {Goerbig}},
  \bibinfo {author} {\bibfnamefont {F.}~\bibnamefont {Guinea}}, \bibinfo
  {author} {\bibfnamefont {G.}~\bibnamefont {Montambaux}}, \ and\ \bibinfo
  {author} {\bibfnamefont {A.~C.}\ \bibnamefont {Neto}},\ }\href@noop {}
  {\bibfield  {journal} {\bibinfo  {journal} {Physical Review B}\ }\textbf
  {\bibinfo {volume} {84}},\ \bibinfo {pages} {045436} (\bibinfo {year}
  {2011})}\BibitemShut {NoStop}%
\bibitem [{\citenamefont {Castro}\ \emph {et~al.}(2007)\citenamefont {Castro},
  \citenamefont {Novoselov}, \citenamefont {Morozov}, \citenamefont {Peres},
  \citenamefont {dos Santos}, \citenamefont {Nilsson}, \citenamefont {Guinea},
  \citenamefont {Geim},\ and\ \citenamefont {Neto}}]{castro_biased_2007}%
  \BibitemOpen
  \bibfield  {author} {\bibinfo {author} {\bibfnamefont {E.~V.}\ \bibnamefont
  {Castro}}, \bibinfo {author} {\bibfnamefont {K.~S.}\ \bibnamefont
  {Novoselov}}, \bibinfo {author} {\bibfnamefont {S.~V.}\ \bibnamefont
  {Morozov}}, \bibinfo {author} {\bibfnamefont {N.~M.~R.}\ \bibnamefont
  {Peres}}, \bibinfo {author} {\bibfnamefont {J.~M. B.~L.}\ \bibnamefont {dos
  Santos}}, \bibinfo {author} {\bibfnamefont {J.}~\bibnamefont {Nilsson}},
  \bibinfo {author} {\bibfnamefont {F.}~\bibnamefont {Guinea}}, \bibinfo
  {author} {\bibfnamefont {A.~K.}\ \bibnamefont {Geim}}, \ and\ \bibinfo
  {author} {\bibfnamefont {A.~H.~C.}\ \bibnamefont {Neto}},\ }\href {\doibase
  10.1103/PhysRevLett.99.216802} {\bibfield  {journal} {\bibinfo  {journal}
  {Phys. Rev. Lett.}\ }\textbf {\bibinfo {volume} {99}},\ \bibinfo {pages}
  {216802} (\bibinfo {year} {2007})}\BibitemShut {NoStop}%
\bibitem [{\citenamefont {Oostinga}\ \emph {et~al.}(2008)\citenamefont
  {Oostinga}, \citenamefont {Heersche}, \citenamefont {Liu}, \citenamefont
  {Morpurgo},\ and\ \citenamefont {Vandersypen}}]{Oostinga2008}%
  \BibitemOpen
  \bibfield  {author} {\bibinfo {author} {\bibfnamefont {J.~B.}\ \bibnamefont
  {Oostinga}}, \bibinfo {author} {\bibfnamefont {H.~B.}\ \bibnamefont
  {Heersche}}, \bibinfo {author} {\bibfnamefont {X.}~\bibnamefont {Liu}},
  \bibinfo {author} {\bibfnamefont {A.~F.}\ \bibnamefont {Morpurgo}}, \ and\
  \bibinfo {author} {\bibfnamefont {L.~M.~K.}\ \bibnamefont {Vandersypen}},\
  }\href {\doibase 10.1038/nmat2082} {\bibfield  {journal} {\bibinfo  {journal}
  {Nature Materials}\ }\textbf {\bibinfo {volume} {7}},\ \bibinfo {pages} {151}
  (\bibinfo {year} {2008})}\BibitemShut {NoStop}%
\bibitem [{\citenamefont {{McCann}}(2006)}]{mccann_asymmetry_2006}%
  \BibitemOpen
  \bibfield  {author} {\bibinfo {author} {\bibfnamefont {E.}~\bibnamefont
  {{McCann}}},\ }\href {\doibase 10.1103/PhysRevB.74.161403} {\bibfield
  {journal} {\bibinfo  {journal} {Physical Review B}\ }\textbf {\bibinfo
  {volume} {74}},\ \bibinfo {pages} {161403} (\bibinfo {year}
  {2006})}\BibitemShut {NoStop}%
\bibitem [{\citenamefont {Min}\ \emph {et~al.}(2007)\citenamefont {Min},
  \citenamefont {Sahu}, \citenamefont {Banerjee},\ and\ \citenamefont
  {MacDonald}}]{min2007}%
  \BibitemOpen
  \bibfield  {author} {\bibinfo {author} {\bibfnamefont {H.}~\bibnamefont
  {Min}}, \bibinfo {author} {\bibfnamefont {B.}~\bibnamefont {Sahu}}, \bibinfo
  {author} {\bibfnamefont {S.}~\bibnamefont {Banerjee}}, \ and\ \bibinfo
  {author} {\bibfnamefont {A.~H.}\ \bibnamefont {MacDonald}},\ }\href {\doibase
  10.1103/PhysRevB.75.155115} {\bibfield  {journal} {\bibinfo  {journal} {Phys.
  Rev. B}\ }\textbf {\bibinfo {volume} {75}},\ \bibinfo {pages} {155115}
  (\bibinfo {year} {2007})}\BibitemShut {NoStop}%
\bibitem [{\citenamefont {{McCann}}\ \emph {et~al.}(2007)\citenamefont
  {{McCann}}, \citenamefont {Abergel},\ and\ \citenamefont
  {Fal’ko}}]{mccann2007}%
  \BibitemOpen
  \bibfield  {author} {\bibinfo {author} {\bibfnamefont {E.}~\bibnamefont
  {{McCann}}}, \bibinfo {author} {\bibfnamefont {D.}~\bibnamefont {Abergel}}, \
  and\ \bibinfo {author} {\bibfnamefont {V.}~\bibnamefont {Fal’ko}},\ }\href
  {\doibase 10.1016/j.ssc.2007.03.054} {\bibfield  {journal} {\bibinfo
  {journal} {Solid State Communications}\ }\textbf {\bibinfo {volume} {143}},\
  \bibinfo {pages} {110} (\bibinfo {year} {2007})}\BibitemShut {NoStop}%
\bibitem [{\citenamefont {Mucha-Kruczy{\'n}ski}\ \emph
  {et~al.}(2009)\citenamefont {Mucha-Kruczy{\'n}ski}, \citenamefont
  {{McCann}},\ and\ \citenamefont
  {Fal’ko}}]{mucha-kruczynski_influence_2009}%
  \BibitemOpen
  \bibfield  {author} {\bibinfo {author} {\bibfnamefont {M.}~\bibnamefont
  {Mucha-Kruczy{\'n}ski}}, \bibinfo {author} {\bibfnamefont {E.}~\bibnamefont
  {{McCann}}}, \ and\ \bibinfo {author} {\bibfnamefont {V.~I.}\ \bibnamefont
  {Fal’ko}},\ }\href {\doibase 10.1016/j.ssc.2009.02.057} {\bibfield
  {journal} {\bibinfo  {journal} {Solid State Communications}\ }\textbf
  {\bibinfo {volume} {149}},\ \bibinfo {pages} {1111} (\bibinfo {year}
  {2009})}\BibitemShut {NoStop}%
\bibitem [{Note1()}]{Note1}%
  \BibitemOpen
  \bibinfo {note} {PVA: Polyvinyl alcohol; PMMA: Poly(methyl
  methacrylate)}\BibitemShut {NoStop}%
\bibitem [{\citenamefont {Lin}\ \emph {et~al.}(2012)\citenamefont {Lin},
  \citenamefont {Lu}, \citenamefont {Yeh}, \citenamefont {Jin}, \citenamefont
  {Suenaga},\ and\ \citenamefont {Chiu}}]{lin2012}%
  \BibitemOpen
  \bibfield  {author} {\bibinfo {author} {\bibfnamefont {Y.-C.}\ \bibnamefont
  {Lin}}, \bibinfo {author} {\bibfnamefont {C.-C.}\ \bibnamefont {Lu}},
  \bibinfo {author} {\bibfnamefont {C.-H.}\ \bibnamefont {Yeh}}, \bibinfo
  {author} {\bibfnamefont {C.}~\bibnamefont {Jin}}, \bibinfo {author}
  {\bibfnamefont {K.}~\bibnamefont {Suenaga}}, \ and\ \bibinfo {author}
  {\bibfnamefont {P.-W.}\ \bibnamefont {Chiu}},\ }\href {\doibase
  10.1021/nl203733r} {\bibfield  {journal} {\bibinfo  {journal} {Nano Letters}\
  }\textbf {\bibinfo {volume} {12}},\ \bibinfo {pages} {414} (\bibinfo {year}
  {2012})}\BibitemShut {NoStop}%
\bibitem [{\citenamefont {Moser}\ \emph {et~al.}(2007)\citenamefont {Moser},
  \citenamefont {Barreiro},\ and\ \citenamefont {Bachtold}}]{moser2007}%
  \BibitemOpen
  \bibfield  {author} {\bibinfo {author} {\bibfnamefont {J.}~\bibnamefont
  {Moser}}, \bibinfo {author} {\bibfnamefont {A.}~\bibnamefont {Barreiro}}, \
  and\ \bibinfo {author} {\bibfnamefont {A.}~\bibnamefont {Bachtold}},\ }\href
  {\doibase http://dx.doi.org/10.1063/1.2789673} {\bibfield  {journal}
  {\bibinfo  {journal} {Applied Physics Letters}\ }\textbf {\bibinfo {volume}
  {91}} (\bibinfo {year} {2007}),\
  http://dx.doi.org/10.1063/1.2789673}\BibitemShut {NoStop}%
\bibitem [{\citenamefont {Goossens}\ \emph {et~al.}(2012)\citenamefont
  {Goossens}, \citenamefont {Calado}, \citenamefont {Barreiro}, \citenamefont
  {Watanabe}, \citenamefont {Taniguchi},\ and\ \citenamefont
  {Vandersypen}}]{goossens2012}%
  \BibitemOpen
  \bibfield  {author} {\bibinfo {author} {\bibfnamefont {A.~M.}\ \bibnamefont
  {Goossens}}, \bibinfo {author} {\bibfnamefont {V.~E.}\ \bibnamefont
  {Calado}}, \bibinfo {author} {\bibfnamefont {A.}~\bibnamefont {Barreiro}},
  \bibinfo {author} {\bibfnamefont {K.}~\bibnamefont {Watanabe}}, \bibinfo
  {author} {\bibfnamefont {T.}~\bibnamefont {Taniguchi}}, \ and\ \bibinfo
  {author} {\bibfnamefont {L.~M.~K.}\ \bibnamefont {Vandersypen}},\ }\href
  {\doibase 10.1063/1.3685504} {\bibfield  {journal} {\bibinfo  {journal}
  {Applied Physics Letters}\ }\textbf {\bibinfo {volume} {100}},\ \bibinfo
  {pages} {073110} (\bibinfo {year} {2012})}\BibitemShut {NoStop}%
\bibitem [{\citenamefont {Ohta}\ \emph {et~al.}(2006)\citenamefont {Ohta},
  \citenamefont {Bostwick}, \citenamefont {Seyller}, \citenamefont {Horn},\
  and\ \citenamefont {Rotenberg}}]{ohta2006}%
  \BibitemOpen
  \bibfield  {author} {\bibinfo {author} {\bibfnamefont {T.}~\bibnamefont
  {Ohta}}, \bibinfo {author} {\bibfnamefont {A.}~\bibnamefont {Bostwick}},
  \bibinfo {author} {\bibfnamefont {T.}~\bibnamefont {Seyller}}, \bibinfo
  {author} {\bibfnamefont {K.}~\bibnamefont {Horn}}, \ and\ \bibinfo {author}
  {\bibfnamefont {E.}~\bibnamefont {Rotenberg}},\ }\href {\doibase
  10.1126/science.1130681} {\bibfield  {journal} {\bibinfo  {journal}
  {Science}\ }\textbf {\bibinfo {volume} {313}},\ \bibinfo {pages} {951}
  (\bibinfo {year} {2006})}\BibitemShut {NoStop}%
\bibitem [{\citenamefont {Russo}\ \emph {et~al.}(2009)\citenamefont {Russo},
  \citenamefont {Craciun}, \citenamefont {Yamamoto}, \citenamefont {Tarucha},\
  and\ \citenamefont {Morpurgo}}]{Russo2009}%
  \BibitemOpen
  \bibfield  {author} {\bibinfo {author} {\bibfnamefont {S.}~\bibnamefont
  {Russo}}, \bibinfo {author} {\bibfnamefont {M.~F.}\ \bibnamefont {Craciun}},
  \bibinfo {author} {\bibfnamefont {M.}~\bibnamefont {Yamamoto}}, \bibinfo
  {author} {\bibfnamefont {S.}~\bibnamefont {Tarucha}}, \ and\ \bibinfo
  {author} {\bibfnamefont {A.~F.}\ \bibnamefont {Morpurgo}},\ }\href {\doibase
  10.1088/1367-2630/11/9/095018} {\bibfield  {journal} {\bibinfo  {journal}
  {New Journal of Physics}\ }\textbf {\bibinfo {volume} {11}},\ \bibinfo
  {pages} {095018} (\bibinfo {year} {2009})}\BibitemShut {NoStop}%
\bibitem [{\citenamefont {Weitz}\ \emph {et~al.}(2010)\citenamefont {Weitz},
  \citenamefont {Allen}, \citenamefont {Feldman}, \citenamefont {Martin},\ and\
  \citenamefont {Yacoby}}]{Weitz2010}%
  \BibitemOpen
  \bibfield  {author} {\bibinfo {author} {\bibfnamefont {R.~T.}\ \bibnamefont
  {Weitz}}, \bibinfo {author} {\bibfnamefont {M.~T.}\ \bibnamefont {Allen}},
  \bibinfo {author} {\bibfnamefont {B.~E.}\ \bibnamefont {Feldman}}, \bibinfo
  {author} {\bibfnamefont {J.}~\bibnamefont {Martin}}, \ and\ \bibinfo {author}
  {\bibfnamefont {A.}~\bibnamefont {Yacoby}},\ }\href {\doibase
  10.1126/science.1194988} {\bibfield  {journal} {\bibinfo  {journal}
  {Science}\ }\textbf {\bibinfo {volume} {330}},\ \bibinfo {pages} {812}
  (\bibinfo {year} {2010})}\BibitemShut {NoStop}%
\bibitem [{\citenamefont {Taychatanapat}\ and\ \citenamefont
  {Jarillo-Herrero}(2010)}]{Thiti2010}%
  \BibitemOpen
  \bibfield  {author} {\bibinfo {author} {\bibfnamefont {T.}~\bibnamefont
  {Taychatanapat}}\ and\ \bibinfo {author} {\bibfnamefont {P.}~\bibnamefont
  {Jarillo-Herrero}},\ }\href {\doibase 10.1103/PhysRevLett.105.166601}
  {\bibfield  {journal} {\bibinfo  {journal} {Phys. Rev. Lett.}\ }\textbf
  {\bibinfo {volume} {105}},\ \bibinfo {pages} {166601} (\bibinfo {year}
  {2010})}\BibitemShut {NoStop}%
\bibitem [{\citenamefont {Varlet}\ \emph
  {et~al.}(2014{\natexlab{b}})\citenamefont {Varlet}, \citenamefont {Liu},
  \citenamefont {Krueckl}, \citenamefont {Bischoff}, \citenamefont {Simonet},
  \citenamefont {Watanabe}, \citenamefont {Richter}, \citenamefont {Ensslin},\
  and\ \citenamefont {Ihn}}]{varlet_fabry_2014}%
  \BibitemOpen
  \bibfield  {author} {\bibinfo {author} {\bibfnamefont {A.}~\bibnamefont
  {Varlet}}, \bibinfo {author} {\bibfnamefont {M.-H.}\ \bibnamefont {Liu}},
  \bibinfo {author} {\bibfnamefont {V.}~\bibnamefont {Krueckl}}, \bibinfo
  {author} {\bibfnamefont {D.}~\bibnamefont {Bischoff}}, \bibinfo {author}
  {\bibfnamefont {P.}~\bibnamefont {Simonet}}, \bibinfo {author} {\bibfnamefont
  {T.~T.}\ \bibnamefont {Watanabe}, \bibfnamefont {K.}}, \bibinfo {author}
  {\bibfnamefont {K.}~\bibnamefont {Richter}}, \bibinfo {author} {\bibfnamefont
  {K.}~\bibnamefont {Ensslin}}, \ and\ \bibinfo {author} {\bibfnamefont
  {T.}~\bibnamefont {Ihn}},\ }\href {\doibase 10.1103/PhysRevLett.113.116601}
  {\bibfield  {journal} {\bibinfo  {journal} {Phys. Rev. Lett.}\ }\textbf
  {\bibinfo {volume} {113}},\ \bibinfo {pages} {116601} (\bibinfo {year}
  {2014}{\natexlab{b}})}\BibitemShut {NoStop}%
\bibitem [{\citenamefont {Williams}\ \emph {et~al.}(2007)\citenamefont
  {Williams}, \citenamefont {{DiCarlo}},\ and\ \citenamefont
  {Marcus}}]{williams_quantum_2007}%
  \BibitemOpen
  \bibfield  {author} {\bibinfo {author} {\bibfnamefont {J.~R.}\ \bibnamefont
  {Williams}}, \bibinfo {author} {\bibfnamefont {L.}~\bibnamefont {{DiCarlo}}},
  \ and\ \bibinfo {author} {\bibfnamefont {C.~M.}\ \bibnamefont {Marcus}},\
  }\href {\doibase 10.1126/science.1144657} {\bibfield  {journal} {\bibinfo
  {journal} {Science}\ }\textbf {\bibinfo {volume} {317}},\ \bibinfo {pages}
  {638} (\bibinfo {year} {2007})}\BibitemShut {NoStop}%
\bibitem [{\citenamefont {\"Ozyilmaz}\ \emph {et~al.}(2007)\citenamefont
  {\"Ozyilmaz}, \citenamefont {Jarillo-Herrero}, \citenamefont {Efetov},
  \citenamefont {Abanin}, \citenamefont {Levitov},\ and\ \citenamefont
  {Kim}}]{ozyilmaz_electronic_2007}%
  \BibitemOpen
  \bibfield  {author} {\bibinfo {author} {\bibfnamefont {B.}~\bibnamefont
  {\"Ozyilmaz}}, \bibinfo {author} {\bibfnamefont {P.}~\bibnamefont
  {Jarillo-Herrero}}, \bibinfo {author} {\bibfnamefont {D.}~\bibnamefont
  {Efetov}}, \bibinfo {author} {\bibfnamefont {D.~A.}\ \bibnamefont {Abanin}},
  \bibinfo {author} {\bibfnamefont {L.~S.}\ \bibnamefont {Levitov}}, \ and\
  \bibinfo {author} {\bibfnamefont {P.}~\bibnamefont {Kim}},\ }\href {\doibase
  10.1103/PhysRevLett.99.166804} {\bibfield  {journal} {\bibinfo  {journal}
  {Phys. Rev. Lett.}\ }\textbf {\bibinfo {volume} {99}},\ \bibinfo {pages}
  {166804} (\bibinfo {year} {2007})}\BibitemShut {NoStop}%
\bibitem [{\citenamefont {Jing}\ \emph {et~al.}(2010)\citenamefont {Jing},
  \citenamefont {Velasco~Jr.}, \citenamefont {Kratz}, \citenamefont {Liu},
  \citenamefont {Bao}, \citenamefont {Bockrath},\ and\ \citenamefont
  {Lau}}]{jing_quantum_2010}%
  \BibitemOpen
  \bibfield  {author} {\bibinfo {author} {\bibfnamefont {L.}~\bibnamefont
  {Jing}}, \bibinfo {author} {\bibfnamefont {J.}~\bibnamefont {Velasco~Jr.}},
  \bibinfo {author} {\bibfnamefont {P.}~\bibnamefont {Kratz}}, \bibinfo
  {author} {\bibfnamefont {G.}~\bibnamefont {Liu}}, \bibinfo {author}
  {\bibfnamefont {W.}~\bibnamefont {Bao}}, \bibinfo {author} {\bibfnamefont
  {M.}~\bibnamefont {Bockrath}}, \ and\ \bibinfo {author} {\bibfnamefont
  {C.~N.}\ \bibnamefont {Lau}},\ }\href {\doibase 10.1021/nl101901g} {\bibfield
   {journal} {\bibinfo  {journal} {Nano Lett.}\ }\textbf {\bibinfo {volume}
  {10}},\ \bibinfo {pages} {4000} (\bibinfo {year} {2010})}\BibitemShut
  {NoStop}%
\bibitem [{\citenamefont {Amet}\ \emph {et~al.}(2013)\citenamefont {Amet},
  \citenamefont {Wiliams}, \citenamefont {Watanabe}, \citenamefont
  {Taniguchi},\ and\ \citenamefont {Goldhaber-Gordon}}]{amet_gate_2013}%
  \BibitemOpen
  \bibfield  {author} {\bibinfo {author} {\bibfnamefont {F.}~\bibnamefont
  {Amet}}, \bibinfo {author} {\bibfnamefont {J.~R.}\ \bibnamefont {Wiliams}},
  \bibinfo {author} {\bibfnamefont {K.}~\bibnamefont {Watanabe}}, \bibinfo
  {author} {\bibfnamefont {T.}~\bibnamefont {Taniguchi}}, \ and\ \bibinfo
  {author} {\bibfnamefont {D.}~\bibnamefont {Goldhaber-Gordon}},\ }\href
  {http://arxiv.org/abs/1307.4408} {\bibfield  {journal} {\bibinfo  {journal}
  {{arXiv:1307.4408} [cond-mat]}\ } (\bibinfo {year} {2013})}\BibitemShut
  {NoStop}%
\bibitem [{\citenamefont {Mucha-Kruczyński}\ \emph {et~al.}(2009)\citenamefont
  {Mucha-Kruczyński}, \citenamefont {Abergel}, \citenamefont {McCann},\ and\
  \citenamefont {Fal’ko}}]{Marcin_2009}%
  \BibitemOpen
  \bibfield  {author} {\bibinfo {author} {\bibfnamefont {M.}~\bibnamefont
  {Mucha-Kruczyński}}, \bibinfo {author} {\bibfnamefont {D.~S.~L.}\
  \bibnamefont {Abergel}}, \bibinfo {author} {\bibfnamefont {E.}~\bibnamefont
  {McCann}}, \ and\ \bibinfo {author} {\bibfnamefont {V.~I.}\ \bibnamefont
  {Fal’ko}},\ }\href {http://stacks.iop.org/0953-8984/21/i=34/a=344206}
  {\bibfield  {journal} {\bibinfo  {journal} {Journal of Physics: Condensed
  Matter}\ }\textbf {\bibinfo {volume} {21}},\ \bibinfo {pages} {344206}
  (\bibinfo {year} {2009})}\BibitemShut {NoStop}%
\end{thebibliography}%

\appendix
\section{Bilayer graphene in quantizing magnetic field}
\label{AppA}

In the presence of an external, perpendicular magnetic field $\vect{B}=(0,0,B)=\nabla\times\vect{A}$, we introduce canonical momentum $\vect{p}=-i\nabla+e\vect{A}$ taking into account the magnetic vector potential $\vect{A}$. In the Landau gauge, $\vect{A}=(0,Bx,0)$, operators $\op{\pi}$ and $\op{\pi}^{\dag}$ in the Hamiltonian in Eq.~\eqref{eqn:4x4hamiltonian} become lowering/raising operators for the Landau functions $\psi_{n}$, \cite{mccann_landau-level_2006}
\begin{align}\begin{split}\label{S1}
& \op{\pi}^{\dag}\psi_{n} = i\frac{\hbar}{\lambda_{B}}\sqrt{2(n+1)}\psi_{n+1}, \\
& \op{\pi}\psi_{n} = -i\frac{\hbar}{\lambda_{B}}\sqrt{2n}\psi_{n-1}, \\
& \psi_{n}=e^{iqy}\frac{1}{\sqrt{2^{n}n!\,\lambda_{B}\sqrt{\pi}}}\exp\!\left(-\frac{1}{2\lambda_{B}^{2}}(x-q\lambda_{B}^{2})^{2}\right)\mathcal{H}_{n}\!\!\left(\frac{x}{\lambda_{b}}-q\lambda_{B}\right), \\
& \lambda_{B}=\sqrt{\frac{\hbar}{eB}},
\end{split}\end{align}
where $\mathcal{H}_{n}(x)$ is the $n$-th Hermite polynomial. Using these relations, one can notice that in the absence of the skew interlayer coupling $\gamma_{3}$ ($v_{3}=0$), the eigenstate of the Hamiltonian, Eq.~\eqref{eqn:4x4hamiltonian}, corresponding to the $n$-th Landau level, can be written in the form $(c_{1}\psi_{n},c_{2}\psi_{n-2},c_{3}\psi_{n-1},c_{4}\psi_{n-1})^{T}$, where $c_{i}$, $i=1,2,3,4$, are some complex numbers (for the case of $n=0,1$, it is enough to put 0 in place of any $\psi_{m<0}$). The coupling $\gamma_{3}$ mixes Landau levels $n$ and $n-3$, so that the wave function component on each sublattice becomes a linear combination of an infinite number of Landau functions $\psi_{n}$. Our procedure to take $\gamma_{3}$ into account in the calculation of the Landau levels follows earlier work which investigated the influence of magnetic fields on the band structure of graphite \cite{inoue_jpsj_1962}. We determine the dispersion by coupling $n_{\mathrm{max}}$ Landau levels (for the spectra shown in this work, $n_{\mathrm{max}}\geq 300$), so that for, e.g., sublattice $B2$ in valley K, we consider $n_{\mathrm{max}}-1$ Landau functions ($n=0,\dots,n_{\mathrm{max}}-2$). Arranging the Landau functions in the order $(\psi_{0}^{A1},\psi_{1}^{A1},\psi_{0}^{A2},\psi_{0}^{B1},\psi_{2}^{A1},\psi_{0}^{B2},\psi_{1}^{A2},\psi_{1}^{B1},\dots)^{T}$ in the valley $K$ and $(\psi_{0}^{B2}, \psi_{1}^{B2}, \psi_{0}^{B1}, -\psi_{0}^{A2}, \psi_{2}^{B2}, -\psi_{0}^{A1}, \psi_{1}^{B1},\\%...
-\psi_{1}^{A2},\dots)^{T}$ in $K'$ (where we used the subscripts to explicitly denote the sublattice associated with a given $\psi_{n}$), and calculating the respective matrix elements, we obtain the following matrix $\op{H}_{\mathrm{num}}$ (in which we already included the interlayer asymmetry $u$ and the strain-induced term $w$) which needs to be diagonalized numerically
\begin{align}\begin{split}
& \op{H}_{\mathrm{num}}=\begin{bmatrix}
\op{H}_{1} & \op{S} & \op{W}_{1} & \op{W}_{2} & 0 & \cdots & 0 \\
\op{S}^{\dag} & \op{H}_{2} & 0 & \op{S} & \op{W}_{3} & \cdots & 0 \\\
\op{W}_{1}^{\dag} & 0 & \op{H}_{3} & 0 & \op{S} & \ddots & \vdots \\
\op{W}_{2}^{\dag} & \op{S}^{\dag} &0 & \op{H}_{4} & 0 &\cdots & \op{W}_{n_{\mathrm{max}}-2} \\
0 & \op{W}_{3}^{\dag} & \op{S}^{\dag} & 0 & \op{H}_{5} & \ddots & \op{S} \\
\vdots & \vdots & \ddots & \ddots & \ddots & \ddots & 0 \\
0 & 0 & \cdots & \op{W}_{n_{\mathrm{max}}-2}^{\dag} & \op{S}^{\dag} & 0 & \op{H}_{n_{\mathrm{max}}}
\end{bmatrix}, \\
& \op{S} = \begin{bmatrix}
0 & \xi w & 0 & 0 \\
0 & 0 & 0 & 0 \\
0 & 0 & 0 & 0 \\
0 & 0 & 0 & 0
\end{bmatrix},\,\,\,\,
\op{H}_{1} = \begin{bmatrix}
\xi\frac{u}{2} & 0 & 0 & 0 \\
0 & \xi\frac{u}{2} & 0 & ix\sqrt{2} \\
0 & 0 & -\xi\frac{u}{2} & \xi\gamma_{1} \\
0 & -ix\sqrt{2} & \xi\gamma_{1} & \xi\frac{u}{2}
\end{bmatrix}, \\
& \op{H}_{n>1} = \begin{bmatrix}
\xi\frac{u}{2} & 0 & 0 & ix\sqrt{2n} \\
0 & -\xi\frac{u}{2} & -ix\sqrt{2(n-1)} & 0 \\
0 & ix\sqrt{2(n-1)} & -\xi\frac{u}{2} & 0 \\
-ix\sqrt{2n} & 0 & 0 & \xi\frac{u}{2}
\end{bmatrix}, \\
& \op{W}_{1} = \begin{bmatrix}
0 & -ix_{3}\sqrt{2} & 0 & 0 \\
0 & w & 0 & 0 \\
0 & 0 & 0 & 0 \\
0 & 0 & 0 & 0
\end{bmatrix},\,\,\,\,
\op{W}_{2} = \begin{bmatrix}
0 & 0 & 0 & 0 \\
0 & -ix_{3}2 & 0 & 0 \\
0 & 0 & 0 & 0 \\
0 & 0 & 0 & 0
\end{bmatrix}, \\
& \op{W}_{n\geq 3} = \begin{bmatrix}
0 & -ix_{3}\sqrt{2n} & 0 & 0 \\
0 & 0 & 0 & 0 \\
0 & 0 & 0 & 0 \\
0 & 0 & 0 & 0
\end{bmatrix},\,\,x=v\sqrt{\hbar eB},\,\,x_{3}=v_{3}\sqrt{\hbar eB}.
\end{split}\end{align}

\section{Self-consistent calculation of $u$}
\label{AppB}

\begin{figure}
\centering
\includegraphics[scale=1]{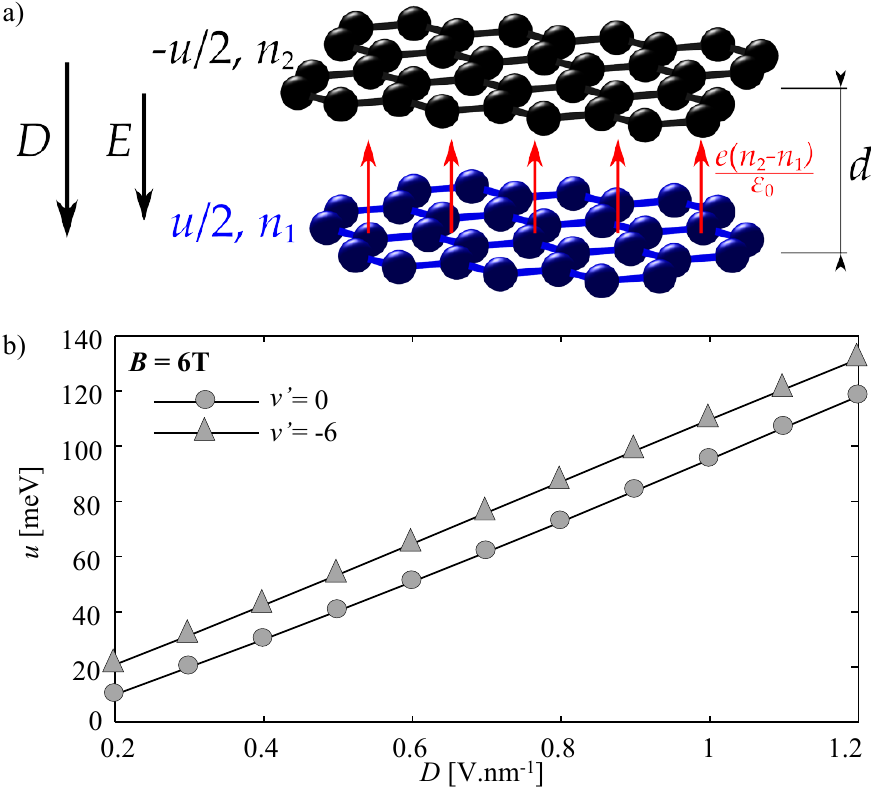}
\caption{(a) Schematic of bilayer graphene with interlayer spacing $d$, in external displacement field $D$ leading to the rearrangement of the densities $n_{1}$ and $n_{2}$ on the bottom and top layer, respectively. This rearrangement screens the external field. (b) Interlayer asymmetry calculated as a function of $D$.}
\label{fig_u}
\end{figure}

In the presence of an external displacement field $D$ applied perpendicularly to the bilayer graphene, electrons minimize their potential energy by rearranging themselves between the graphene layers \cite{mccann_asymmetry_2006,mucha-kruczynski_influence_2009}. As a result, the external field is screened, yielding an effective electric field between the layers of magnitude $E$, as shown in Fig.~\ref{fig_u}(a). The interlayer asymmetry $u$ can be related to this electric field, following

\begin{equation}
\label{eq_u}
u=\epsilon_{1}-\epsilon_{2}= eEd = \frac{ed}{\epsilon_{0}} [D+e(n_{1} - n_{2})],
\end{equation}

where $\epsilon_{1}$ ($\epsilon_{2}$) is the on-site energy on the first (second) layer, $e$ is the absolute value of electron charge, $d$ is the interlayer spacing, $\epsilon_{0}$ is the permittivity of vacuum, and $n_{1}$ ($n_{2}$) is the charge density on the first (second) layer. Because the charge densities $n_{1}$ and $n_{2}$ depend on the details of the band structure, which is, in turn, influenced by $u$, Eq.~\eqref{eq_u} needs to be solved self-consistently.

In the presence of a magnetic field, the difference ($n_{1}-n_{2}$) can be determined using the capacity of a single (spin degenerate) Landau level, $2eB/h$, where $h$ is the Planck constant, and the knowledge of the electronic eigenstates as obtained in \ref{AppA}. As explained there, in the presence of the trigonal warping, the component $c_{n}^{i}$ of the electron wave function of the $n$-th Landau level on sublattice $i$ is a linear combination of the Landau level functions, $c_{n}^{i}=\sum_{\substack{j}} c_{n,j}^{i} \psi_{j}$. The charge density difference between the layers is then

\begin{equation}
\label{eq_density}
(n_{1}-n_{2}) = 2 \frac{eB}{h} \sum_{\substack{n,j}} (|c_{n,j}^{A1}|^{2} + |c_{n,j}^{B1}|^{2} - |c_{n,j}^{A2}|^{2} - |c_{n,j}^{B2}|^{2}),
\end{equation}

with the summation over $n$ taking into account only those Landau levels that are filled for a specified filling factor $\nu'$. In the numerical calculation, we assume an initial value of $u_{in}$, which determines through Eq.~\eqref{eq_u} and \eqref{eq_density} the output asymmetry $u_{out}$. We then search for $u_{in}$ such that the self-consistent value of $u$, fulfilling the condition $u_{out}$ = $u_{in}$, is found. The number of the eigenvalues included in the summation in Eq.~\eqref{eq_density} needs to be varied with magnetic field, so that big enough range of the bilayer spectrum, $\sim 1~\rm{eV}$ away from the neutrality point, is included in the calculation. Because the calculation is based on the single-particle Hamiltonian [Eq.~\eqref{hamilt-with-u-full}] and the spin degeneracy is implied, we are limited to even filling factors. At the same time, experimental features described in the main text are related to filling factors $\nu' = -1$ to $\nu' = -6$. Hence, we show in Fig.~\ref{fig_u}(b) the interlayer asymmetry $u$ as a function of the displacement field $D$ for filling factors $\nu' = 0$ and $\nu' = -6$, to provide estimates on the range of $u$ to be expected in the experiment. Note that the Landau levels at the edge of the band, originating in the zero-energy state of gapless BLG, are localised dominantly on one of the layers \cite{mccann_landau-level_2006,Marcin_2009}. As a result, their contribution to one of the charge densities $n_{1}/n_{2}$ is weakly dependent on $u$, and significant compared to the contribution of deeper Landau levels, shared more equally between both layers. At small displacement fields $D$, this leads to large charge density that is placed on one layer only and cannot be redistributed through the self-consistent scheme, resulting in unphysical results and lack of convergence. Hence, in Fig.~\ref{fig_u}(b) we do not extend the displacement axis to zero. Our calculation also neglects exchange effects, what is only justified for large $u$.

\end{document}